\documentclass[10pt,journal]{IEEEtran}
\IEEEoverridecommandlockouts %
\makeatletter
\newcommand{\KeepWithNext}[1]{%
  \par\penalty-100\begingroup
  \setlength{\dimen@}{#1}%
  \dimen@ii\pagegoal\advance\dimen@ii-\pagetotal
  \ifdim\dimen@>\dimen@ii\break\fi
  \endgroup
}
\makeatother

\usepackage{blkarray}                                      %
\usepackage{capt-of}
\usepackage{placeins}
\usepackage{stfloats}[2025/06/18]
\usepackage{graphicx}                                      %
\usepackage{amsmath}
\usepackage{amssymb}
\usepackage{amsfonts}
\usepackage{amsthm}
\usepackage[mathcal]{eucal}
\usepackage{mathrsfs}
\usepackage{etoolbox}
\usepackage{booktabs}
\AtBeginEnvironment{table}{\renewcommand{\arraystretch}{1.14}}
\AtBeginEnvironment{table*}{\renewcommand{\arraystretch}{1.14}}
\usepackage{enumerate}
\usepackage{multirow}
\usepackage[subrefformat=parens,farskip=0pt,justification=centering]{subfig}
\usepackage{caption}
\usepackage{color}
\usepackage[nocompress]{cite}                              %
\usepackage{comment}                                       %
\usepackage{soul}                                          %
\soulregister\cite7
\soulregister\ref7
\soulregister\pageref7
\usepackage{etoolbox}                                      %
\usepackage{url}
\usepackage{nth}                                           %
\usepackage{bm}                                            %
\usepackage{courier}
\usepackage{threeparttable}
\usepackage{xcolor,colortbl}
\usepackage{footnote}
\usepackage{listings}
\usepackage[skins,breakable,listings]{tcolorbox}

\usepackage{setspace}                                      %
\usepackage[inline]{enumitem}
\usepackage{verbatim}
\usepackage[bookmarks=false]{hyperref}
\usepackage{orcidlink}
\hypersetup{
    colorlinks=true,
    citecolor=blue,
    linkcolor=blue,
    urlcolor=blue,
    pdftitle   = {RTLREX: LLM-Guided Framework for Exact High-Level RTL Recovery},
    pdfauthor  = {Wuqian Tang, Fangzhou Liu, Ruijie Wang, TingTing Hwang, Bei Yu, and Chun-Yao Wang},
    pdfsubject = {Logic-to-RTL reverse engineering},
    pdfkeywords = {circuit reverse engineering, RTL recovery, formal equivalence, logic netlists, truth tables, representation quality},
}

\usepackage{tikz}
\usetikzlibrary{patterns}
\usetikzlibrary{positioning,calc,fit,decorations.pathmorphing,shapes.geometric, shapes.gates.logic.US, calc}
\usetikzlibrary{arrows,arrows.meta,decorations.markings,shapes,shapes.arrows}
\usetikzlibrary{decorations,decorations.pathreplacing}
\usetikzlibrary{backgrounds}
\usepackage{pgfplots}
\usepackage{pgfplotstable}
\usepgfplotslibrary{groupplots}
\usepackage{scalefnt}
\pgfplotsset{compat=newest}
\usepackage{pifont}                                        %
\usepackage{cleveref}
\Crefname{equation}{Equation}{Equations}
\crefname{appendix}{Appendix}{Appendixes}
\Crefname{appendix}{Appendix}{Appendixes}
\Crefformat{figure}{Fig.~#2#1#3}                           %
\Crefname{subfigure}{Fig.}{Figs.}
\Crefname{figure}{Fig.}{Figs.}
\Crefformat{table}{TABLE~#2#1#3}                           %
\usepackage[figuresright]{rotating}

\definecolor{CUHKorange}{RGB}{244,106,18} %
\definecolor{CUHKblue}{RGB}{0,111,190}    %
\definecolor{CUHKgreen}{RGB}{0,127,128}   %
\definecolor{CUHKred}{RGB}{228,46,36}     %
\definecolor{CUHKyellow}{RGB}{198,148,34} %
\definecolor{CUHKdark}{RGB}{114,44,114}   %
\definecolor{CUHKmiddle}{RGB}{144,44,144} %
\definecolor{CUHKlight}{RGB}{167,44,167} 
\definecolor{CUHKpurple}{RGB}{117,15,109}
\definecolor{CUHKgold}{RGB}{221,163,0}
\definecolor{CUHKribbon}{RGB}{244,223,176}
\definecolor{CUHKblack}{RGB}{34,24,21}
\definecolor{RECOVERblue}{RGB}{28,91,148}
\definecolor{RECOVERorange}{RGB}{190,92,25}
\definecolor{RECOVERgreen}{RGB}{35,118,92}

\renewcommand{\bf}[1]{\textbf{#1}}
\renewcommand{\tt}[1]{\texttt{#1}}

\newcommand{\minisection}[1]{\vspace{.1in}\noindent{\textbf{#1}}.}

\usepackage{tcolorbox}
\tcbuselibrary{skins,breakable}
    {\endtcolorbox}
    {\endtcolorbox}

\patchcmd{\subsection}{3.5ex plus 1.5ex minus 1.5ex}
  {3.5ex plus 0.3ex minus 1.5ex}{}{
  \PackageError{rtlrex}{Subsection spacing patch failed}{Check the IEEEtran version.}}
\patchcmd{\subsection}{0.7ex plus .5ex minus 0ex}
  {0.7ex plus .1ex minus 0ex}{}{
  \PackageError{rtlrex}{Subsection spacing patch failed}{Check the IEEEtran version.}}

\newtheorem{myexample}{\textbf{Example}}

\crefname{cecproposition}{Proposition}{Propositions}
\Crefname{cecproposition}{Proposition}{Propositions}

\crefname{mytheorem}{Theorem}{Theorems}
\crefname{mylemma}{Lemma}{Lemmas}
\crefname{myclaim}{Claim}{Claims}
\crefname{myproperty}{Property}{Properties}
\crefname{mycorollary}{Corollary}{Corollaries}

\RequirePackage[normalem]{ulem} %
\RequirePackage{color}\definecolor{RED}{rgb}{1,0,0}\definecolor{BLUE}{rgb}{0,0,1} %

\graphicspath{{./figs/}{../}}

\begin{document}
\bstctlcite{rtlrexIEEEReferenceStyle}

\title{
RTLREX: LLM-Guided Framework for\\
Exact High-Level RTL Recovery
}
\author{Wuqian Tang~\orcidlink{0009-0008-5042-5062},~\IEEEmembership{Student Member,~IEEE},
Fangzhou Liu~\orcidlink{0009-0008-9370-4504},
Ruijie Wang~\orcidlink{0009-0000-7966-6259},
TingTing Hwang~\orcidlink{0000-0002-7206-560X},~\IEEEmembership{Member,~IEEE},
Bei Yu~\orcidlink{0000-0001-6406-4810},~\IEEEmembership{Senior Member,~IEEE}, and
Chun-Yao Wang~\orcidlink{0000-0002-4536-214X},~\IEEEmembership{Senior Member,~IEEE}%
\thanks{This work was supported in part by the National Science and Technology
Council (NSTC) under Grants NSTC 112-2218-E-007-014, NSTC 112-2221-E-007-106-MY2,
NSTC 112-2221-E-007-108, NSTC 113-2640-E-011-003, NSTC 113-2221-E-007-082-MY3,
NSTC 114-2221-E-007-125-MY3, NSTC 114-2218-E-007-002, and
NSTC 114-2221-E-007-126-MY3. \textit{(Corresponding author: Wuqian Tang.)}}%
\thanks{Wuqian Tang, Ruijie Wang, TingTing Hwang, and Chun-Yao Wang are with the
Department of Computer Science, National Tsing Hua University, Hsinchu 300044, Taiwan
(e-mail: wqtang@cs.nthu.edu.tw; wrj651121@gmail.com;
tingting@cs.nthu.edu.tw; wcyao@cs.nthu.edu.tw).}%
\thanks{Fangzhou Liu and Bei Yu are with the Department of Computer
Science and Engineering, The Chinese University of Hong Kong, Shatin, N.T.,
Hong Kong SAR (e-mail: fzliu23@cse.cuhk.edu.hk; byu@cse.cuhk.edu.hk).}}

\maketitle
\thispagestyle{plain}
\pagestyle{plain}
\newcommand{\RecoverCaseRows}{%
\textit{test01} & 35 & \textbf{2} & \textbf{2} & 2 & 0 & 0 & \textbf{2} \\
\textit{test02} & 58 & \textbf{2} & \textbf{2} & 1 & 1 & 0 & \textbf{2} \\
\textit{test03} & 313 & \textbf{3} & \textbf{3} & 3 & 0 & 0 & \textbf{3} \\
\textit{test04} & 7416 & \textbf{6} & \textbf{6} & 4 & 1 & 0 & \textbf{6} \\
\textit{test05} & 1290 & \textbf{6} & \textbf{6} & 3 & 3 & 0 & \textbf{6} \\
\addlinespace[2pt]
\textit{test06} & 4959 & \textbf{3} & \textbf{3} & 2 & 1 & 0 & \textbf{3} \\
\textit{test07} & 2845 & 9 & 8 & 0 & 1 & 3 & \textbf{6} \\
\textit{test08} & 2696 & \textbf{4} & \textbf{4} & 2 & 0 & 0 & \textbf{4} \\
\textit{test09} & 7250 & \textbf{3} & \textbf{3} & 1 & 1 & 0 & \textbf{3} \\
\textit{test10} & 2658 & \textbf{4} & \textbf{4} & 2 & 1 & 0 & \textbf{4} \\
\addlinespace[2pt]
\textit{test11} & 1795 & \textbf{2} & \textbf{2} & 0 & 1 & 0 & \textbf{2} \\
\textit{test12} & 11227 & \textbf{23} & \textbf{23} & 11 & 12 & 0 & \textbf{23} \\
\textit{test13} & 931 & 18 & \textbf{17} & 2 & 0 & 9 & \textbf{17} \\
\textit{test14} & 66 & \textbf{2} & \textbf{2} & 0 & 0 & 0 & \textbf{2} \\
\textit{test15} & 3468 & 12 & \textbf{9} & 4 & 2 & 3 & \textbf{9} \\
\addlinespace[2pt]
\textit{test16} & 721 & \textbf{5} & \textbf{5} & 0 & 0 & 0 & \textbf{5} \\
\textit{test17} & 938 & \textbf{7} & \textbf{7} & 2 & 0 & 0 & \textbf{7} \\
\textit{test18} & 96 & 12 & 7 & 3 & 0 & 0 & \textbf{6} \\
\textit{test19} & 1981 & 91 & 50 & 5 & 1 & 43 & \textbf{49} \\
\textit{test20} & 4836 & \textbf{18} & \textbf{18} & 6 & 6 & 6 & \textbf{18} \\
\addlinespace[2pt]
\textit{test21} & 1620 & 59 & 49 & 28 & 0 & 1 & \textbf{31} \\
\textit{test22} & 91584 & Failed & \textbf{25} & 12 & 13 & 0 & \textbf{25} \\
\textit{test23} & 299 & Failed & \textbf{3} & 2 & 0 & 1 & \textbf{3} \\
\textit{test24} & 15245 & Failed & Timeout & 50 & 24 & 0 & \textbf{86} \\
\textit{test25} & 215 & Failed & 20 & 0 & 0 & 0 & \textbf{19} \\
\addlinespace[2pt]
\textit{test26} & 16098 & Failed & Timeout & 13 & 12 & 2 & \textbf{46} \\
\textit{test27} & 3160 & Failed & Timeout & 19 & 0 & 25 & \textbf{49} \\
\textit{test28} & 8538 & Failed & Timeout & 9 & 12 & 0 & \textbf{27} \\
\textit{test29} & 688 & 53 & 69 & 6 & 1 & 7 & \textbf{17} \\
\textit{test30} & 14371 & 24 & \textbf{18} & 6 & 6 & 6 & \textbf{18} \\
\midrule
Recovered & -- & 23/30 & 26/30 & \multicolumn{4}{c}{\textbf{30/30}} \\
\textbf{Avg. (all 30)} & 6913.23 & -- & -- & 6.60 & 3.30 & 3.53 & 16.60 \\
}

\newcommand{\TableFiveAccepted}{30}
\newcommand{\TableFiveTimeoutAccepted}{0}
\newcommand{\TableFiveCertified}{30}
\newcommand{\OursPACertified}{30}
\newcommand{\OursPAV}{30}
\newcommand{\OursPAC}{30}
\newcommand{\DeterministicPAF}{19}
\newcommand{\DeterministicPAY}{19}
\newcommand{\DeterministicPAC}{17}
\newcommand{\OursDateCommon}{23}
\newcommand{\OursDateLhsCost}{243}
\newcommand{\OursDateRhsCost}{368}
\newcommand{\OursDateWins}{8}
\newcommand{\OursDateTies}{15}
\newcommand{\OursDateLosses}{0}
\newcommand{\OursDateLhsMedian}{6}
\newcommand{\OursDateRhsMedian}{6}
\newcommand{\OursWolfexCommon}{26}
\newcommand{\OursWolfexLhsCost}{290}
\newcommand{\OursWolfexRhsCost}{365}
\newcommand{\OursWolfexLhsMean}{11.15}
\newcommand{\OursWolfexRhsMean}{14.04}
\newcommand{\OursWolfexMeanReduction}{20.5}
\newcommand{\OursWolfexWins}{6}
\newcommand{\OursWolfexTies}{20}
\newcommand{\OursWolfexLosses}{0}
\newcommand{\OursWolfexLhsMedian}{6}
\newcommand{\OursWolfexRhsMedian}{6.50}
\newcommand{\SkillAblationPACommon}{21}
\newcommand{\SkillAblationPALhsCost}{421}
\newcommand{\SkillAblationPARhsCost}{158,190}
\newcommand{\SkillAblationPAWins}{15}
\newcommand{\SkillAblationPATies}{6}
\newcommand{\SkillAblationPALosses}{0}
\newcommand{\SkillAblationPALhsMedian}{17}
\newcommand{\SkillAblationPARhsMedian}{23}
\newcommand{\DeterministicAblationPACommon}{19}
\newcommand{\DeterministicAblationPALhsCost}{273}
\newcommand{\DeterministicAblationPARhsCost}{19,820}
\newcommand{\DeterministicAblationPAWins}{14}
\newcommand{\DeterministicAblationPATies}{5}
\newcommand{\DeterministicAblationPALosses}{0}
\newcommand{\DeterministicAblationPALhsMedian}{6}
\newcommand{\DeterministicAblationPARhsMedian}{18}
\newcommand{\DeterministicAblationTTCommon}{73}
\newcommand{\DeterministicAblationTTLhsCost}{3,778}
\newcommand{\DeterministicAblationTTRhsCost}{3,778}
\newcommand{\DeterministicAblationTTWins}{0}
\newcommand{\DeterministicAblationTTTies}{73}
\newcommand{\DeterministicAblationTTLosses}{0}
\newcommand{\DeterministicAblationTTLhsMedian}{57}
\newcommand{\DeterministicAblationTTRhsMedian}{57}
\newcommand{\OursDateLhsMean}{10.57}
\newcommand{\OursDateRhsMean}{16.00}
\newcommand{\OursDateMeanReduction}{34.0}
\newcommand{\PAFullCount}{30}

\newcommand{\AigCommon}{12}
\newcommand{\AigFrozenNodes}{28,282}
\newcommand{\AigReferenceNodes}{19,310}
\newcommand{\AigLower}{1}
\newcommand{\AigEqual}{3}
\newcommand{\AigHigher}{8}

\newcommand{\TTFamilyEndpointRows}{%
Floating-point unary & 40 & 40/40 & 0/0 \\
Floating-point binary & 10 & 8/8 & 0/0 \\
Floating-point fused multiply--add & 1 & 0/0 & 0/0 \\
Floating-point dot product & 1 & 0/0 & 0/0 \\
Fixed-point dot product & 2 & 0/0 & 0/0 \\
Fixed-point reduction & 1 & 0/0 & 0/0 \\
Integer arithmetic & 25 & 25/25 & 0/0 \\
Unclassified & 20 & 0/0 & 0/0 \\
\textbf{Total} & \textbf{100} & \textbf{73/73} & \textbf{0/0} \\
}

\newcommand{\UpdatedSkillV}{100}
\newcommand{\UpdatedSkillH}{80}
\newcommand{\UpdatedSkillPolicyV}{100}
\newcommand{\UpdatedSkillTimeouts}{1}
\newcommand{\UpdatedSkillMissingFinal}{0}
\newcommand{\UpdatedSkillExitFailures}{1}
\newcommand{\UpdatedSkillUsageAvailable}{99}
\newcommand{\UpdatedSkillUsageMissing}{1}
\newcommand{\UpdatedSkillGenerationMedian}{174.1}
\newcommand{\UpdatedSkillGenerationSum}{17896.3}
\newcommand{\UpdatedSkillVFailures}{0}
\newcommand{\UpdatedSkillPolicyViolations}{0}
\newcommand{\UpdatedSkillInfrastructureErrors}{0}
\newcommand{\RetainedNoSkillV}{81}
\newcommand{\RetainedNoSkillH}{74}
\newcommand{\RetainedNoSkillPolicyV}{78}
\newcommand{\RetainedNoSkillTimeouts}{51}
\newcommand{\RetainedNoSkillMissingFinal}{17}
\newcommand{\RetainedNoSkillExitFailures}{51}
\newcommand{\RetainedNoSkillUsageAvailable}{50}
\newcommand{\RetainedNoSkillUsageMissing}{50}
\newcommand{\RetainedNoSkillGenerationMedian}{600.0}
\newcommand{\RetainedNoSkillGenerationSum}{43548.3}
\newcommand{\RetainedNoSkillVFailures}{19}
\newcommand{\RetainedNoSkillPolicyViolations}{3}
\newcommand{\RetainedNoSkillInfrastructureErrors}{0}
\newcommand{\UpdatedCostVCommon}{81}
\newcommand{\UpdatedCostVWithWins}{66}
\newcommand{\UpdatedCostVTies}{14}
\newcommand{\UpdatedCostVNoWins}{1}
\newcommand{\UpdatedCostVRatioN}{81}
\newcommand{\UpdatedCostVMedianRatio}{0.778}
\newcommand{\UpdatedCostVMeanReduction}{95.7}
\newcommand{\UpdatedRuntimeVWithSkillMedian}{160.3}
\newcommand{\UpdatedCostVWithSkillMean}{53.81}
\newcommand{\UpdatedCostVWithSkillMedian}{23.00}
\newcommand{\UpdatedRuntimeVNoSkillMedian}{501.6}
\newcommand{\UpdatedCostVNoSkillMean}{1253.78}
\newcommand{\UpdatedCostVNoSkillMedian}{28.00}
\newcommand{\UpdatedCostHCommon}{74}
\newcommand{\UpdatedCostHWithWins}{60}
\newcommand{\UpdatedCostHTies}{14}
\newcommand{\UpdatedCostHNoWins}{0}
\newcommand{\UpdatedCostHRatioN}{74}
\newcommand{\UpdatedCostHMedianRatio}{0.798}
\newcommand{\UpdatedCostHMeanReduction}{31.1}
\newcommand{\UpdatedRuntimeHWithSkillMedian}{155.9}
\newcommand{\UpdatedCostHWithSkillMean}{21.66}
\newcommand{\UpdatedCostHWithSkillMedian}{21.00}
\newcommand{\UpdatedRuntimeHNoSkillMedian}{456.3}
\newcommand{\UpdatedCostHNoSkillMean}{31.42}
\newcommand{\UpdatedCostHNoSkillMedian}{26.00}
\newcommand{\UpdatedCostRawVCommon}{81}
\newcommand{\UpdatedCostPolicyExcluded}{0}
\newcommand{\UpdatedSkillRetainedCoverageRows}{%
RTLREX & 100 & 100/100 & 80/100 \\
LLM agent baseline & 100 & 81/100 & 74/100 \\
}
\newcommand{\UpdatedSkillRetainedCostRows}{%
$V$ & 81 & 66 & 14 & 1 & 95.7\% \\
$H$ & 74 & 60 & 14 & 0 & 31.1\% \\
}

\newcommand{\UpdatedSkillGenerationMean}{179.0}
\newcommand{\RetainedNoSkillGenerationMean}{435.5}
\newcommand{\UpdatedGenerationMeanReduction}{58.9}
\newcommand{\UpdatedRuntimeVWithSkillMean}{141.9}
\newcommand{\UpdatedRuntimeVNoSkillMean}{396.9}
\newcommand{\UpdatedRuntimeVMeanReduction}{64.2}
\newcommand{\UpdatedRuntimeHWithSkillMean}{133.2}
\newcommand{\UpdatedRuntimeHNoSkillMean}{378.0}
\newcommand{\UpdatedRuntimeHMeanReduction}{64.8}

\begin{abstract}
Hardware designs may be available only as gate-level netlists or
input--output descriptions when the original register-transfer-level (RTL)
design and documentation are not accessible. Low-level representations obscure word-level computation and its finite-width behavior. Recovering high-level RTL from such representations therefore helps engineers understand a circuit's functionality and enables RTL rewriting and resynthesis. A plausible function candidate is also insufficient unless the generated RTL is equivalent to the original circuit for all input combinations. In this paper, we present RTLREX, a large language model
(LLM)-guided framework that coordinates domain-specific tools for function
recovery, RTL generation, independent verification, and cost-guided
rewriting. Independent verification establishes equivalence between the
recovered RTL and the original circuit. The LLM uses tool feedback to select further analyses, revise RTL candidates,
and propose syntax-cost reductions. Experiments on 30 benchmarks
from Problem A of the ICCAD 2022 CAD Contest show that RTLREX recovers
equivalent high-level RTL for all benchmarks, surpassing the state-of-the-art
method WolFEx in recovery coverage. On benchmarks recovered by both methods,
RTLREX reduces mean syntax cost by \OursWolfexMeanReduction\% relative to WolFEx.
On complete truth-table benchmarks from the IWLS 2026 Contest, our framework reduces
mean syntax cost by 95.7\% relative to an LLM agent baseline on the benchmarks
validated by both methods.
\end{abstract}


\section{Introduction}\label{sec:intro}

\IEEEPARstart{M}{odern} digital systems are built by integrating and reusing
hardware components~\cite{li2012behavioral}. However, the original
register-transfer-level (RTL) descriptions and design documentation may be
unavailable, particularly for third-party components and legacy
hardware~\cite{subramanyan2014reverse,narayanan2023wordmodels}. Engineers
must therefore infer circuit functionality from available representations,
such as gate-level netlists or input--output behavior, to validate existing
designs, diagnose problems, and adapt implementations to new requirements.

Recovering high-level functionality from these representations is
challenging because the word-level computation is not explicit. When
generating a gate-level netlist, logic synthesis preserves functionality but
can obscure the original arithmetic and control structure as it transforms RTL
into Boolean gates. Even a simple arithmetic expression may therefore be
distributed across many Boolean
operations~\cite{li2013wordrev,albartus2020dana}. A complete truth table
similarly specifies every input--output pair without identifying the
word-level expressions behind those values. Circuit reverse engineering
addresses this challenge by obtaining an understandable and functionally
equivalent high-level RTL description.

\begin{figure}[!t]
\centering
\includegraphics[width=\columnwidth]{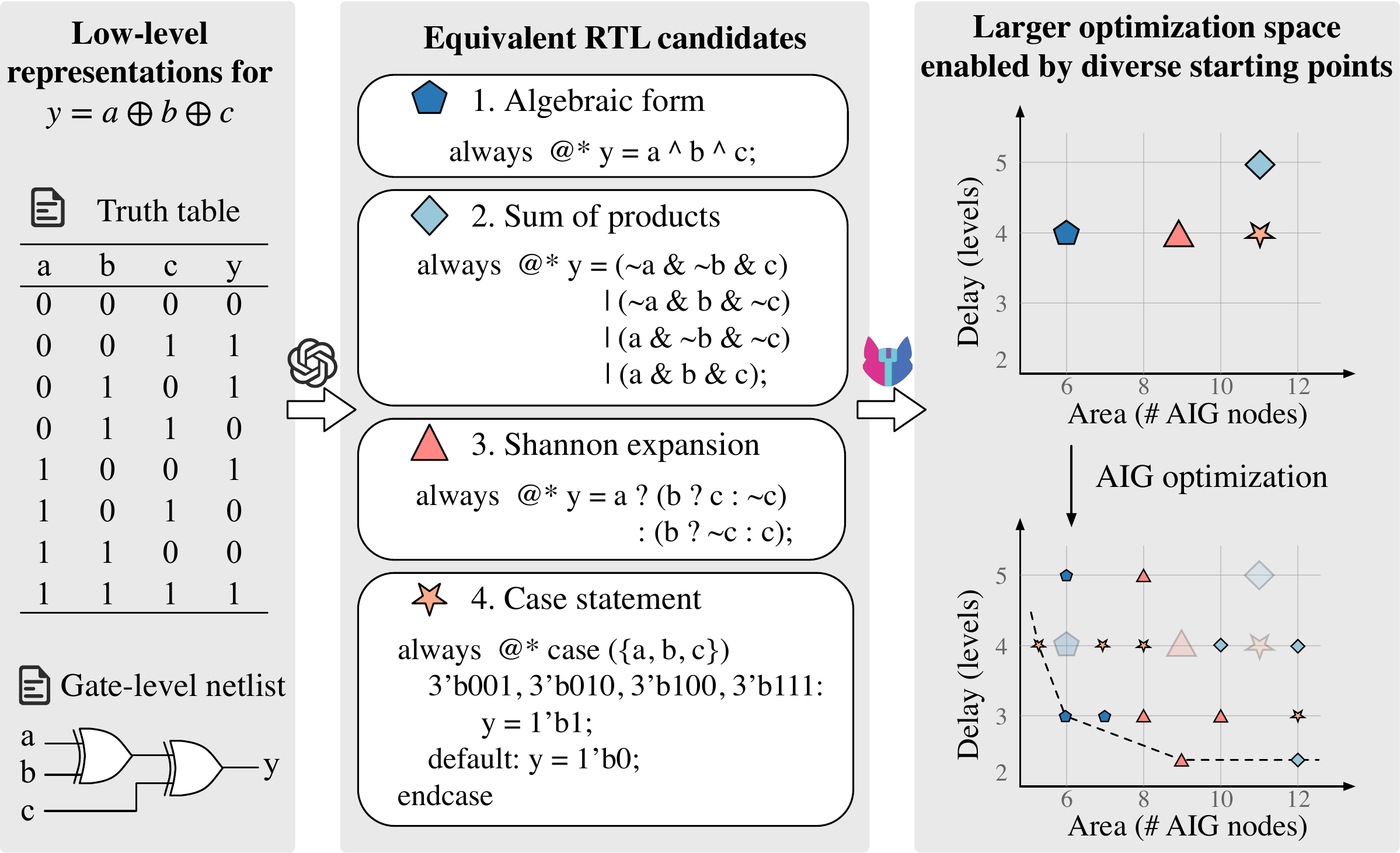}
\caption{High-level RTL recovery enables rewriting and resynthesis.
The area--delay plots illustrate potential tradeoffs schematically.
AIG stands for and-inverter graph.}
\label{fig:motivation}
\end{figure}

The recovered high-level RTL makes the underlying computation
explicit, facilitating functional analysis and RTL rewriting.
As illustrated in \Cref{fig:motivation}, a function represented
by a truth table or gate-level netlist can be expressed in
multiple equivalent RTL forms, including an algebraic expression,
a sum-of-products, a Shannon expansion, and a case statement.
Although these descriptions implement the same function, they
expose different logic structures and may yield
distinct area--delay tradeoffs, as shown in the upper-right plot.
Starting from these different logic structures, subsequent AIG
optimization can explore additional alternatives,
potentially reaching better area--delay tradeoffs, as illustrated
in the lower-right plot. Thus, a single RTL representation may
not expose the full optimization potential of the underlying
function. These potential benefits motivate our focus on
recovering compact, equivalent RTL designs and evaluating their downstream
utility through arithmetic-aware resynthesis and technology
mapping.

Prior work on recovering high-level representations from gate-level
netlists includes arithmetic-function recovery, word-level structure
recovery, and netlist-to-RTL generation.
Hybrid-DATE~\cite{tang2024hybrid} and WolFEx~\cite{ho2023wolfex}
recover arithmetic functionality from gate-level netlists, but their
reported recovery coverage remains incomplete on the ICCAD 2022 CAD
Contest Problem A benchmarks.
ReBERT~\cite{zhang2025rebert} identifies word-level groupings
without reconstructing complete RTL functionality.
Learning-based approaches, including
Gate-Breaker~\cite{faruque2025gatebreaker} and
GenEDA~\cite{fang2025geneda}, explore direct netlist-to-RTL
generation, although Gate-Breaker reports lower testbench pass rates
for more complex benchmarks. Despite these
advances, reliably recovering compact, high-level RTL
across diverse circuit functions remains challenging.
Rule-based function recovery searches predefined function classes, such as addition,
multiplication, comparisons, and bitwise exclusive OR, under supported
operand and control configurations. When these searches find no function candidate
matching the observed outputs, engineers
must further interpret circuit observations, select additional analyses,
and revise function candidates. Prior tools likewise support manual
investigation by exposing word-level structure and accepting user-specified
operation sequences~\cite{li2013wordrev}. A large language model (LLM) can guide this exploration
by relating tool feedback to function candidates, requesting targeted
analyses, and constructing or revising RTL. This motivates combining
LLM-guided exploration with executable recovery tools and independent
verification.

Beyond recovering high-level functionality, establishing
exact equivalence to the original circuit remains a critical
challenge. Reported recovery success is not always supported
by a conclusive equivalence proof. For example, GenEDA
evaluates generated RTL through testbench simulation, which
cannot guarantee correctness for untested input combinations.
In its reported implementation, WolFEx returns \texttt{TRUE} when its
combinational equivalence checking (CEC) procedure finds no counterexample
within the time limit~\cite{ho2023wolfex}. A timeout without a
counterexample, however, does not establish equivalence.
Moreover, multiplication-rich
netlists pose difficulties for conventional bit-level
equivalence checking~\cite{mahzoon2019revsca}. These challenges motivate coupling
RTL recovery with independent, arithmetic-aware equivalence
verification and correctness-preserving RTL rewriting.

We present RTLREX, an LLM-guided framework for exact
high-level RTL recovery from gate-level netlists or complete truth tables. Guided by
domain-specific instructions, the LLM coordinates automatic function
recognition and feedback-driven revision of function and RTL candidates. It also uses syntax-cost
feedback to propose compact equivalent RTL expressions. The
domain-specific procedures and tools are packaged as an Agent
Skill~\cite{agentskills2026format}.
RTLREX uses recovered
arithmetic expressions and independently verified intermediate circuits to
establish netlist equivalence. For complete truth tables, independent RTL
evaluation confirms matching outputs for every input combination.

Compared with non-LLM methods such as Hybrid-DATE~\cite{tang2024hybrid}
and WolFEx~\cite{ho2023wolfex}, RTLREX recovers
more benchmarks and achieves lower mean syntax cost on the benchmarks
jointly recovered with each method. These results support integrating LLM
guidance with domain-specific tools.

The main contributions of this paper are as follows:
\begin{itemize}[leftmargin=*,nosep]
\item We propose RTLREX, a framework whose LLM uses circuit observations,
verification feedback, and syntax cost to select analyses and construct or
revise compact RTL candidates from gate-level netlists or complete truth tables.
\item We develop an equivalence-checking workflow that uses recovered arithmetic
expressions and independently verified intermediate circuits to compose
netlist proofs. The workflow also validates truth-table recovery over the
complete input domain.
\item We develop a cost-guided RTL rewriting procedure in which the LLM uses
syntax-cost feedback to simplify arithmetic expressions and control conditions.
Independent verification and cost evaluation ensure that each adopted rewrite
preserves equivalence and lowers syntax cost.
\item We evaluate RTLREX on gate-level netlists and complete truth tables,
demonstrating exact RTL recovery and lower mean syntax cost than the other
methods on common successful benchmarks. We also demonstrate the benefits of recovered arithmetic structure
for resynthesis and technology mapping.
\end{itemize}

\section{Preliminaries}\label{sec:prelim}

\subsection{Related Work}
\label{sec:background}

Prior work on gate-level reverse engineering spans structural analysis,
behavioral inference, and arithmetic reasoning. Structural and behavioral
approaches recover word boundaries, datapaths, and functional blocks from
flattened logic
\cite{li2013wordrev,subramanyan2014reverse,li2012behavioral,
soeken2015simulation,yu2016automatic,he2021graph}.
However, identifying these components alone is insufficient for complete
RTL recovery: the operations and control logic that produce the outputs
must also be reconstructed with correct bit widths, signedness, and
truncation. Separately, arithmetic verification methods encode gate
behavior as polynomials and reduce output differences to establish
arithmetic identities, with specialized techniques for multiplier circuits
\cite{yu2016formal,mahzoon2019revsca,mahzoon2022revsca2,chen2026reveal}.
While these methods can establish arithmetic identities, proving such
relations does not by itself recover a complete RTL description.

To recover explicit arithmetic expressions, Hybrid-DATE~\cite{tang2024hybrid}
combines structural analysis with black-box testing.
It first analyzes the functional support of output bits and their
exclusive-sum-of-products (XSOP) representations to identify arithmetic
expressions with unknown coefficients. Input dependencies are further
examined to infer conditional operations and their control conditions.
Random simulation then generates input--output equations, and Gaussian
elimination determines the unknown coefficients. This process reconstructs
compact arithmetic RTL, but relies on the expression forms inferred
from the circuit structure.

WolFEx~\cite{ho2023wolfex} addresses the same recovery task through
multiple complementary extraction techniques rather than relying
primarily on structurally inferred expression templates.
It first orders output words according to their functional dependencies
and applies three extraction strategies. Polynomial rewriting derives
word-level subexpressions from algebraic representations of the gate-level
logic. Symbolic regression searches for function candidates from
input--output samples and iteratively refines them using counterexamples
from combinational equivalence checking (CEC). Linear coefficient
fitting provides another approach for recovering expressions with
unknown coefficients. After extracting function candidates and determining
operand signedness, WolFEx employs e-graph rewriting and integer linear programming (ILP)-based
extraction to minimize expression cost while exploiting common
subexpression sharing. However, its published implementation may return
a positive CEC result when no counterexample is found within the time
limit, which does not necessarily establish equivalence.

Beyond these domain-specific recovery techniques, LLMs have been
explored for RTL generation, repair, and electronic design automation (EDA)
tool orchestration.
RTLCoder~\cite{liu2025rtlcoder} trains a specialized model for
natural-language-to-RTL generation. MAGE~\cite{zhao2025mage}
coordinates RTL generation and debugging through testbench feedback,
while AutoChip and RTLFixer use compilation or simulation feedback
to refine generated RTL
\cite{thakur2023autochip,tsai2024rtlfixer}.
ChatEDA~\cite{wu2024chateda} uses LLM planning to coordinate EDA
tools. However, generating or repairing RTL from specifications
differs from recovering functionality from a low-level circuit.

More directly targeting gate-level reverse engineering,
ReBERT~\cite{zhang2025rebert} uses a fine-tuned Bidirectional Encoder
Representations from Transformers (BERT) model to identify word-level
groupings, which reveal operand structures
without reconstructing complete RTL functionality.
Gate-Breaker~\cite{faruque2025gatebreaker} investigates
LLM-based netlist-to-RTL generation.
GenEDA~\cite{fang2025geneda} aligns a circuit encoder with
an LLM decoder to generate high-level RTL from gate-level
netlists. It evaluates the generated RTL through testbench
simulation, which checks correctness only for the tested
input combinations rather than establishing equivalence
over the complete input domain.

RTLREX combines LLM-guided use of domain-specific tools with independent
equivalence checking to recover and refine high-level RTL.

\subsection{Problem Formulation}\label{sec:problem}

Given a combinational gate-level netlist or a complete truth table $O$,
the task is to recover a synthesizable high-level RTL description $R$
with the same input--output interface. Let $x$ denote an $n$-bit input
vector, and let $f_O$ and $f_R$ denote the output functions of the original
representation and the recovered RTL. We consider binary inputs, with each
input bit taking a value of 0 or 1. Equivalence requires
\begin{equation}
f_R(x)=f_O(x),\qquad\forall x\in\{0,1\}^{n}
\label{eq:functional_exact}
\end{equation}
Thus, every pair of corresponding output bits must be identical under every
input combination. RTLREX also checks that the generated RTL can be
processed by Yosys and has no structural errors, such as undriven outputs.
Equivalence checking returns one of three outcomes:
\begin{itemize}
    \item $\mathrm{EQ}$ (equivalent): Equivalence is established.
    \item $\mathrm{NEQ}$ (not equivalent): A counterexample is found.
    \item $\mathrm{UNK}$ (unknown): Checking is inconclusive, including a timeout.
\end{itemize}
Moreover, RTLREX uses the syntax cost described in \Cref{sec:syntax_cost} to guide recovery
toward compact high-level expressions and reports the lowest-cost verified
RTL candidate found. Validated bit-level RTL remains eligible for the coverage
metric defined in \Cref{sec:metrics}.

\setcounter{topnumber}{1}
\section{RTLREX Framework}\label{sec:method}

\subsection{Overview}\label{sec:method_overview}
\begin{figure}[t]
\centering
\includegraphics[width=\columnwidth]{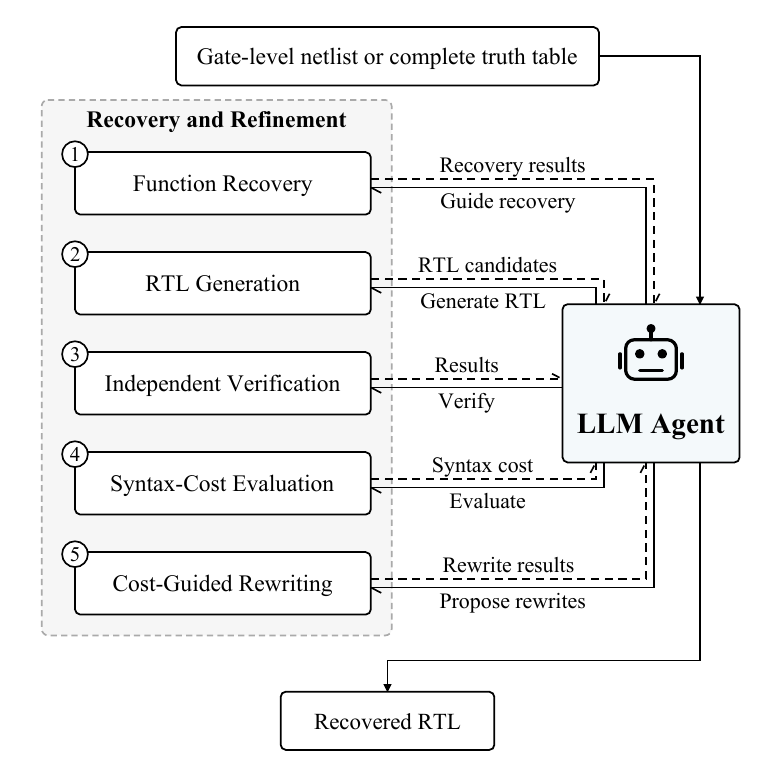}
\caption{Overview of RTLREX.}
\label{fig:overview}
\end{figure}

RTLREX recovers compact high-level RTL from a gate-level netlist or a
complete truth table while preserving the input--output interface and
establishing functional equivalence. As shown in \Cref{fig:overview}, the
LLM coordinates five stages: \textbf{function recovery}, \textbf{RTL
generation}, \textbf{independent verification}, \textbf{syntax-cost
evaluation}, and \textbf{cost-guided rewriting}. The LLM selects the initial
recovery route based on whether the input is a gate-level netlist or a
complete truth table, together with the interface information.
In \Cref{fig:overview}, solid interaction arrows represent actions initiated
by the LLM, and dashed arrows represent results returned by the tools.
When the LLM invokes a tool, the tool executes its processing workflow and
returns the results upon completion. The LLM combines these results with
earlier observations to select the next tool call or construct or revise
function and RTL candidates.
The domain-specific procedures and tools are packaged as an Agent
Skill~\cite{agentskills2026format}. The \hyperref[app:prompts]{Appendix}
describes the accompanying software and task instructions.

In Step~1 (\textbf{Function Recovery}), the recovery tools identify function candidates
by determining operations, operand mappings, and finite-width behavior. If no function candidate
matches the observed outputs, the LLM uses circuit observations to select further
probes and construct or revise expressions. In Step~2 (\textbf{RTL
Generation}), the generator automatically translates supported recovered
functions into RTL. When a recovered function combines operations or control
conditions beyond the generator's supported forms, the LLM constructs the
corresponding RTL and revises RTL candidates using subsequent checking results.

In Step~3 (\textbf{Independent Verification}), the checkers test RTL validity
and establish equivalence to the original representation. The LLM uses
reported RTL errors or equivalence counterexamples to request further
analysis in Step~1 or revise the RTL candidate in Step~2. In
Step~4 (\textbf{Syntax-Cost Evaluation}), the evaluator computes $Q$ for
verified RTL. The LLM examines the RTL, its syntax cost, and the cost breakdown. When a
promising simplification is identified, the LLM proposes a rewrite in Step~5 (\textbf{Cost-Guided Rewriting}). Each rewrite
returns to verification and cost evaluation. A verified RTL candidate with a lower
syntax cost replaces the current best RTL. When refinement ends or the time or rewrite-attempt limit is reached,
RTLREX returns the lowest-cost verified RTL found.

\begin{figure}[t]
\centering
\includegraphics[width=\columnwidth]{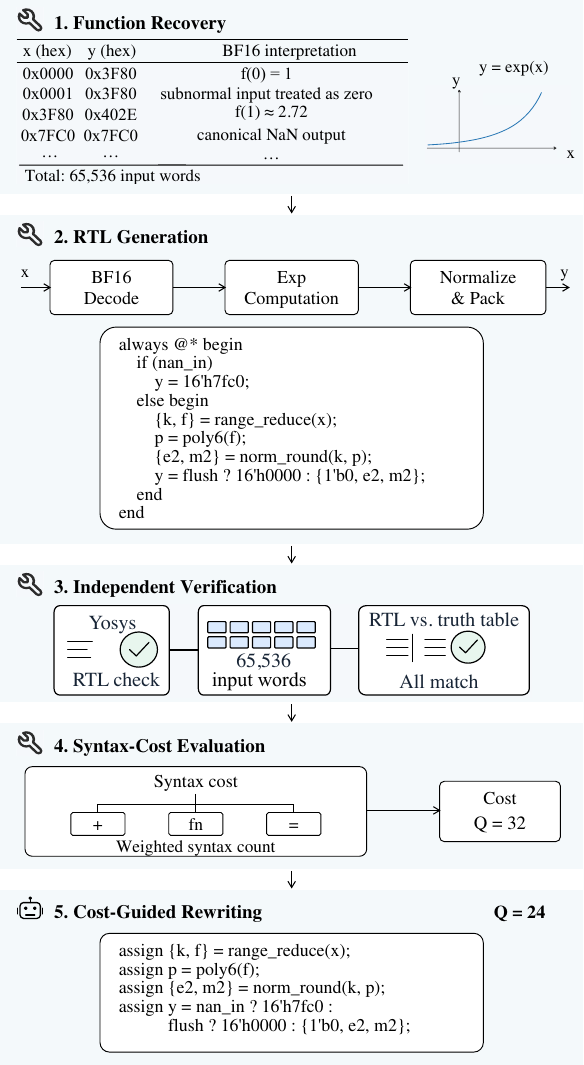}
\caption{RTL recovery and rewriting for \textit{ex200}.
NaN denotes not a number.}
\label{fig:recovery_example}
\end{figure}

\Cref{fig:recovery_example} illustrates recovery for \textit{ex200}.
The tool invoked by the LLM identifies a bfloat16 (BF16) exponential
function matching all 65\,536 input words. Its generator emits arithmetic
RTL: BF16 Decode extracts input fields, Exp Computation performs range
reduction and polynomial evaluation, and Normalize \& Pack normalizes,
rounds, and encodes the result. The RTL also handles special values.
Yosys checks RTL validity, and complete-domain simulation confirms
agreement with the truth table. The initial cost is $Q=32$. From the RTL
and cost feedback, the LLM proposes replacing procedural branches with
conditional expressions. Independent verification confirms equivalence,
and the cost decreases to $Q=24$. The following subsections describe the
five stages.

\begin{table}[!t]
\centering
\caption{Supported Function Classes}
\label{tab:recognizers}
\small
\renewcommand{\arraystretch}{1.12}
\setlength{\tabcolsep}{3pt}
\begin{tabular*}{\columnwidth}{@{\extracolsep{\fill}}>{\raggedright\arraybackslash}p{0.17\columnwidth}>{\raggedright\arraybackslash}p{0.48\columnwidth}>{\raggedright\arraybackslash}p{0.25\columnwidth}@{}}
\toprule
Input & Function classes & Comparison inputs \\
\midrule
Complete truth table & Boolean and affine functions, integer arithmetic,
fixed-point and floating-point operations & All input combinations \\
Gate-level netlist & Arithmetic expressions, comparisons, control-dependent computations &
Targeted and random input vectors \\
\bottomrule
\end{tabular*}
\end{table}

\subsection{Function Recovery}\label{sec:recognition}
The LLM selects recovery operations using the input representation,
interface widths, and the observations returned by analysis tools. For
netlists, these observations include input dependencies and simulated
outputs. For complete truth tables, the tools can compare the predictions of function candidates
with every input combination. Each function candidate
specifies an operation, its operands, and its finite-width behavior.
Recovery can proceed incrementally across individual outputs, internal signal
groups, or control-dependent branches. For netlists, the recovery tools fit
expressions for selected outputs and control conditions. When fitting fails,
the LLM uses dependency analysis and simulation feedback to select further
probes and construct or revise expressions for the unresolved portions.
\Cref{tab:recognizers} summarizes the supported function classes and
comparison inputs.

Building on structural analysis, simulation, and coefficient fitting used
in prior recovery methods~\cite{tang2024hybrid,ho2023wolfex}, RTLREX
implements automatic recognizers for arithmetic expressions, comparisons,
control-dependent computations, and Boolean and numerical functions.
For netlists, dependency analysis traces the fan-in cones of selected signals.
Targeted and random simulations reveal participating operands and their
interactions. Targeted inputs include zero, unit values, operand boundaries, and paired
operand variations.
For a $w$-bit output, arithmetic fitting uses
\begin{equation}
\widehat y(x)=\Bigl(c_0+\sum_i c_i\,t_i(x)\Bigr)\bmod 2^w
\label{eq:modular}
\end{equation}
where $x$ is the input vector, $c_0$ and $c_i$ are integer coefficients,
and $t_i(x)$ are terms such as operands, squares, or operand products.
The modulo operation retains the lowest $w$ bits. The all-zero input gives
the constant offset. Varying one operand distinguishes linear and higher-order
terms, while paired variations reveal product terms through output changes
not explained by individual contributions. For control-dependent behavior,
the tools hold selected control bits fixed, fit a separate expression for
each combination, and combine the expressions into conditional RTL. Boundary
tests and additional random inputs check function candidates for width, signedness,
overflow, and control errors before independent equivalence checking.

\begin{figure}[t]
\centering
\includegraphics[width=0.97\linewidth]{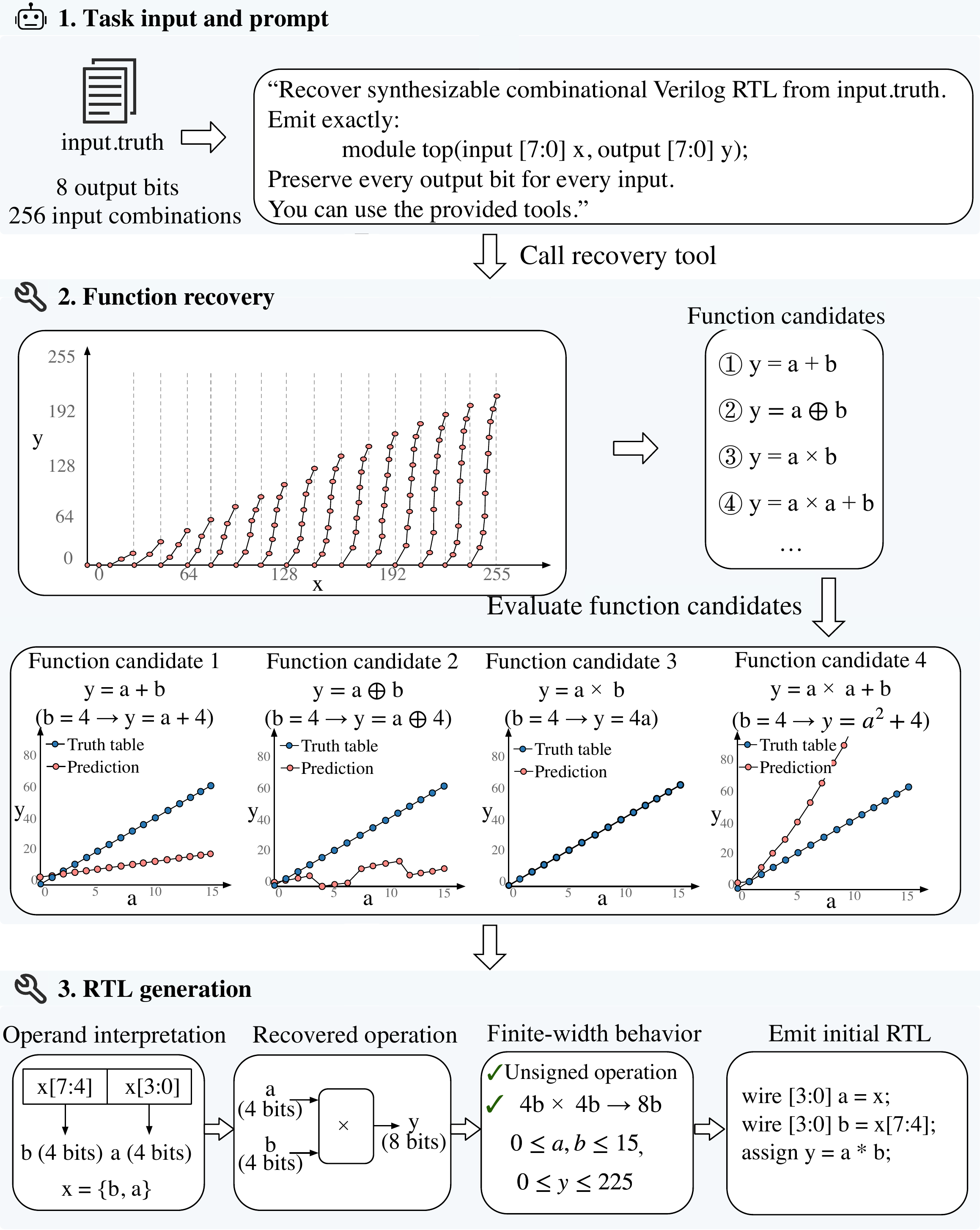}
\caption{Unsigned 4-bit multiplication recovery on \textit{ex255}:
task input and prompt (stage~1), function recovery (stage~2), and RTL
generation (stage~3).}
\label{fig:method_blueprint}
\end{figure}

For complete truth tables, the recognizers examine Boolean and affine
functions, integer arithmetic, and supported fixed-point and floating-point
operations. An affine Boolean model expresses each output bit as the
exclusive OR (XOR) of selected input bits and a constant. The output for
the all-zero input gives the constant bit. XORing this constant with the
output for each one-hot input gives the corresponding coefficient. The tools check the resulting expression against
every table entry. Numerical models specify operand packing, signedness,
and arithmetic widths. Floating-point models additionally specify rounding
and special-value handling for the supported format, including signed zero,
subnormal values, overflow, infinities, and NaN values. Numerical function candidates are
also compared against every table entry.

When automatic recovery finds no function candidate that matches the
observed outputs, the LLM
examines the returned dependencies, simulated outputs, and failed comparisons
of function candidates to determine what to investigate next. It can propose a
conditional expression or revise an interpretation of intermediate widths
and signedness, then request targeted tests of control combinations or
operand boundaries. Comparing predicted and observed outputs distinguishes
these interpretations and guides further expression revisions. A function candidate
consistent with the observations proceeds to RTL generation and independent
verification. Verification errors and available counterexamples can initiate
another round of analysis and correction.

For \textit{test27}, automatic recovery initially produced no word-level
function candidate. The LLM used dependency-analysis and simulation tools to construct
conditional arithmetic expressions. After an intermediate RTL candidate still
failed checking, it requested boundary probes for the sum of operands
\texttt{in13} and \texttt{in24}. A sum of 511 contributed 511, whereas 512
contributed $-512$ in the affected branches, distinguishing a signed 10-bit
intermediate from an unsigned sum. The LLM revised those branches to
sign-extend the 10-bit result to 13 bits. The resulting RTL passed exact
equivalence checking. This illustrates how targeted probing resolves a
specific ambiguity left by automatic recovery.

\Cref{fig:method_blueprint} illustrates recovery for \textit{ex255}.
In Stage~1, the LLM receives a complete truth table containing 256 input
combinations and eight output bits, together with a task prompt specifying
the module interface, synthesizable RTL generation, and exact agreement
with the truth table. Following these instructions, the LLM invokes the
recovery tool in Stage~2. The tool evaluates function candidates with different operations and operand interpretations.
With $a=x[3:0]$ and $b=x[7:4]$, successive 16-input segments vary $a$ while
holding $b$ fixed. Each segment is linear in $a$ with slope $b$, suggesting
$y=ab$. The illustrated slice $b=4$ distinguishes multiplication, $4a$,
from addition, $a+4$, bitwise XOR, $a\oplus4$, and the quadratic model,
$a^2+4$. The four comparison plots overlay each function candidate's
predictions with the truth-table outputs, showing agreement for multiplication
and mismatches for the other illustrated candidates.
The tool confirms unsigned 4-bit multiplication by matching all
256 input combinations, yielding an 8-bit output. The corresponding RTL
generation is described in \Cref{sec:rtl_generation}.

\subsection{RTL Generation}\label{sec:rtl_generation}
RTLREX translates a recovered function into combinational RTL using its
operation, operand mapping, and finite-width behavior. For supported functions, the generator emits declarations and expressions while preserving
the original module interface. Integer functions become word-level expressions,
affine Boolean functions become XOR expressions, and control-dependent functions
use conditional assignments or combinational branches. Intermediate signals
and type conversions encode widths and signedness, and output assignments
implement truncation. Widening uses sign extension for signed intermediates and zero extension
for unsigned intermediates, after computing each result at its recovered
width. Supported numerical functions use operation-specific
generators for field decoding, arithmetic, rounding, and output encoding,
as illustrated by the BF16 exponential function in
\Cref{fig:recovery_example}.

When the recovered expression combines operations or control conditions
beyond the generator's supported forms, the LLM constructs the corresponding
RTL from the recovered function and circuit observations. It specifies
intermediate widths, signedness, and branch conditions explicitly. If RTL
validity checking reports an error or equivalence checking finds a mismatch,
the LLM uses the diagnostic information or available counterexample to
revise the RTL candidate. This connects function recovery to RTL construction:
observations determine the computation, and checking feedback helps resolve
errors in its finite-width implementation.

\begin{figure}[t]
\centering
\tcbinputlisting{enhanced,listing only,
  listing file={data/prompts/ex255_task_excerpt.txt},
  title={Task prompt excerpt},
  colback=black!3,colframe=black!20,colbacktitle=black!3,coltitle=black,
  fonttitle=\normalfont\footnotesize\bfseries,
  boxrule=0.4pt,arc=2pt,left=5pt,right=5pt,top=4pt,bottom=4pt,
  before skip=6pt,after skip=6pt,
  listing options={language={},breaklines=true,breakatwhitespace=true,
    breakautoindent=false,breakindent=0pt,
    basicstyle=\ttfamily\footnotesize,columns=fullflexible,
    keepspaces=true,showstringspaces=false,numbers=none,frame=none,
    aboveskip=0pt,belowskip=0pt}}
\caption{Task prompt excerpt for RTL recovery.}
\label{fig:task_prompt}
\end{figure}

Stage~3 in \Cref{fig:method_blueprint} illustrates tool-assisted
RTL generation for \textit{ex255}. The generator translates the recovered
unsigned multiplication into combinational RTL. \Cref{fig:task_prompt}
provides a longer excerpt of the task prompt summarized in Stage~1 of
\Cref{fig:method_blueprint}.

The generated RTL preserves the 8-bit input $x$ and output $y$
and declares two unsigned 4-bit operands, $a$ and $b$.
Assigning $x$ to $a$ selects its lower 4 bits, while $b$ explicitly
selects $x[7:4]$, preserving the operand interpretation recovered
in Stage~2. The expression $y=a\times b$ implements the identified
multiplication. Since both operands range from 0 to 15, their
maximum product is $15\times15=225$, which fits within the 8-bit
output without truncation.

In this example, the LLM inspects the generated RTL and verification feedback
before returning the final implementation.

Even when a function candidate matches the observations, the generated
RTL can differ because of bit widths, signedness, truncation, or
branch conditions. RTLREX therefore independently checks the
generated RTL's validity and equivalence before evaluating its
cost, as described in \Cref{sec:verification}.

\subsection{Independent Verification}\label{sec:verification}
RTLREX compares each RTL candidate against the original design with matched
port correspondence, widths, and signedness. Direct equivalence checking and
arithmetic verification can be combined by proving that internal netlist
signals equal corresponding RTL expressions. Failing simulation vectors and,
when available, formal counterexamples provide feedback for the LLM to select
further probes and revise RTL candidates. Unsupported or inconclusive results
are reported as $\mathrm{UNK}$.

\minisection{Verification Methods}
RTLREX applies direct equivalence checking or arithmetic verification to
the input representation. For gate-level netlists, Yosys and ABC are used
for AIG-based CEC, whereas Yosys and Z3 are used for
bit-vector satisfiability modulo theories (SMT) checking.
Both methods examine whether there exists an input combination for which
corresponding outputs differ. For complete truth tables, we implement a bit-parallel
evaluator that packs the values of multiple input combinations into bit
vectors and evaluates supported operations using bitwise computation.
Icarus Verilog provides complete-domain simulation when the evaluator
encounters unsupported operations. Both approaches compare every output
bit against the truth table over the complete input domain.

\begin{figure*}[!t]
  \centering
  \includegraphics[width=\textwidth]{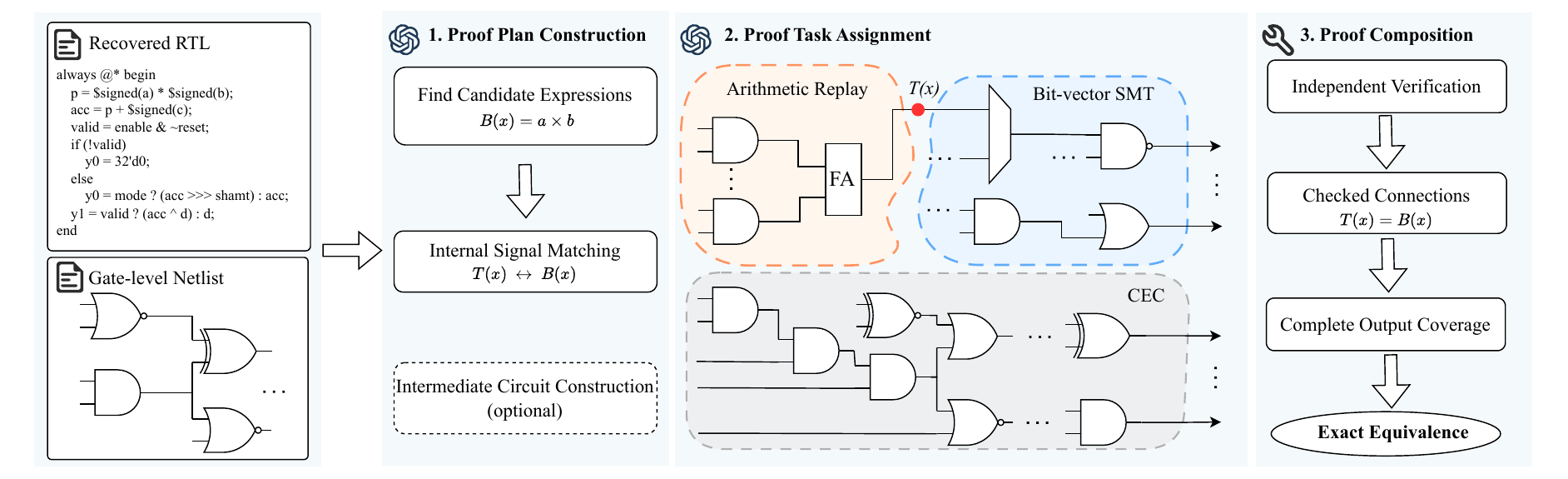}
  \caption{Proof composition in RTLREX. Independently verified relations
  connect local proofs to establish equivalence for all outputs.
  The circuit structures are schematic. FA denotes a full adder.}
  \label{fig:proof_composition}
\end{figure*}

\label{sec:algebraic_proof}
For multiplication-rich netlists that are difficult to verify directly with
CEC or SMT, RTLREX employs arithmetic verification based on algebraic
rewriting~\cite{yu2016formal}. It represents the original netlist and recovered
RTL as integer polynomials and examines whether their difference reduces to
zero modulo $2^W$, where $W$ is the output width. A zero difference establishes
equivalence. For example, a half adder has a sum bit $s=a\oplus b$ and a carry
bit $c=a\land b$. Interpreting Boolean values as integers in $\{0,1\}$ gives
$s=a+b-2ab$ and $c=ab$. The complete 2-bit unsigned output $\{c,s\}$ has the
numeric value $s+2c$ and satisfies
\begin{equation}
s+2c=(a+b-2ab)+2ab=a+b
\label{eq:adder_identity}
\end{equation}
Suppose the original netlist produces this value $Y_O=s+2c$, and the recovered
RTL computes the full 2-bit unsigned result $Y_R=a+b$, including its carry.
By \eqref{eq:adder_identity}, their difference is
\begin{equation}
Y_O-Y_R=(s+2c)-(a+b)=0
\end{equation}
In practice, RTLREX identifies local arithmetic identities by exhaustively
comparing small Boolean subcircuits with up to three boundary inputs.
An independent checker rechecks these identities and applies them to reduce
the output difference. This process is termed \emph{arithmetic replay} in
\Cref{fig:proof_composition}. To limit polynomial growth, RTLREX organizes
the reduction by output-bit significance. Control-dependent logic is checked
separately for every combination of selected control bits, while the other
inputs remain unrestricted. Implementation details are available in the software
companion described in~\hyperref[app:prompts]{Appendix}.

\minisection{Compositional Verification}
When direct checking of a gate-level netlist $O$ against recovered RTL $R$
is inconclusive, RTLREX combines smaller proofs through the three-stage
workflow in \Cref{fig:proof_composition}. Let $x$ denote the input vector,
and let $f_O(x)$ and $f_R(x)$ denote the corresponding output vectors.
The goal is to establish $f_O(x)=f_R(x)$ for every $x$.

\begin{enumerate}[label=\arabic*),leftmargin=*,itemsep=0.5ex,topsep=0.5ex,parsep=0pt]
\item Proof Plan Construction.
The agent proposes intermediate expressions from the recovered RTL and uses
circuit analysis and simulation to find candidate matching signals in the
netlist. In \Cref{fig:proof_composition}, the multiplication assigned to
``p'' gives $B(x)=a\times b$, and $T(x)$ denotes a candidate matching signal.
The equality $T(x)=B(x)$ must be proved before it can be used in another
proof. For difficult arithmetic proofs, the agent may also restructure
selected gate-level logic into an intermediate circuit and then separately
prove its equivalence to the original netlist and the recovered RTL.

\item Proof Task Assignment.
The agent organizes the candidate signal relations and target outputs into
proof tasks and assigns a checking method to each task. In
\Cref{fig:proof_composition}, arithmetic replay checks the multiplication
logic (orange), bit-vector SMT checks downstream word-level operations
(blue), and CEC checks the other logic feeding the outputs (gray).
Recovered RTL supplies the word-level expressions for these checks.
More than one method may apply to a task. For example, CEC can also be used
to verify gate-level arithmetic transformations. Each task specifies the
input--output correspondence, widths, and signedness.

\item Proof Composition.
RTLREX verifies the internal relations before using them in downstream
proof tasks. For the internal signal $T(x)$ marked by the red dot in
\Cref{fig:proof_composition}, RTLREX first proves $T(x)=B(x)$ for all $x$.
It then replaces this internal signal on both sides with an additional
input $t$ and proves $F(x,t)=G(x,t)$ for all values of $x$ and $t$.
Here, $F$ and $G$ describe the remaining downstream logic. The downstream
proof therefore holds when $t$ is set to the verified value $T(x)=B(x)$,
giving
\begin{equation}
F(x,T(x))=G(x,B(x))
\label{eq:free_cut}
\end{equation}
This use of the verified internal equality connects the two proofs, as shown
in the final stage of \Cref{fig:proof_composition}. The proofs also cover the outputs driven by the remaining gate-level
logic, shown in gray in \Cref{fig:proof_composition}. Every original output
bit must be covered by the proofs.
For an intermediate circuit $I$, separately proving $O\equiv I$ and
$I\equiv R$ establishes $O\equiv R$ by transitivity.
\end{enumerate}

The following example illustrates how an intermediate circuit combines
CEC and arithmetic replay.

\begin{myexample}
\label{sec:netlist_proof_example}
Consider \textit{test22} from the 30 gate-level netlists of Problem A in
the ICCAD 2022 CAD Contest (PA-30)~\cite{cadcontest2022problemA}.
Its recovered RTL $R$ expresses a sum of 13 multiplication products,
while the gate-level netlist $O$ contains 830 input bits and 69 output bits.
RTLREX constructs an intermediate gate-level circuit $I$ that preserves
the upstream logic and rewrites the final summation as an equivalent
adder network with an explicit sum-and-carry structure. Since $O$ and $I$
share most of their gate-level structure, CEC can verify $O\equiv I$.
Then arithmetic replay uses the explicit addition structure of $I$ to
prove $I\equiv R$. Combining these independently verified relations gives
$O\equiv R$, as shown in \eqref{eq:intermediate_chain}:
\begin{equation}
  (O\equiv I)\land(I\equiv R)
  \;\Longrightarrow\; O\equiv R
  \label{eq:intermediate_chain}
\end{equation}
\end{myexample}

Composition uses only relations that have already been proved, and the
checker verifies all connections between proof tasks. When tasks are split
by control conditions, they must cover every combination of the selected
control bits, while leaving the remaining inputs unrestricted. All original output bits
must be covered. Together, these requirements ensure that the combined
proof applies to the complete original design.

RTLREX also rejects unknown or undriven outputs. After equivalence is
established, the RTL candidate proceeds to syntax-cost evaluation. Every
subsequent RTL rewrite must pass the same checks again.

\subsection{Syntax-Cost Evaluation}\label{sec:syntax_cost}

RTLREX evaluates the compactness of RTL candidates using syntax cost
$Q$, the weighted sum of charged operators and constructs in an RTL design.
The syntax-cost weights for PA-30 follow the ICCAD 2022 CAD Contest Problem A
specification~\cite{cadcontest2022problemA}.
For the 100 complete truth-table benchmarks from the IWLS 2026 Contest
(TT-100)~\cite{iwls2026contest}, we extend these weights by increasing
the charges for bitwise operators, reduction operators, case items,
and primitive-gate instances,
while assigning zero cost to unary negation and adding zero-cost
SystemVerilog keywords.  These adjustments favor compact word-level
representations. \Cref{tab:cost_profiles} summarizes the syntax-cost weights
for PA-30 and TT-100.

Weights are charged per occurrence unless otherwise specified.
Zero-weight keywords, including ``else'' and ``default'', incur
no separate charge. Submodule costs are included for each instantiation.
The syntax-cost weights for TT-100 are adapted from the PA-30 contest
weights for this work. They are not part of the official IWLS scoring rules.

The evaluator counts operators and constructs in RTL designs,
excluding declaration ranges and event controls, and distinguishes
the relational operator ``\textless='' from nonblocking assignment.
Function bodies are counted as written without expansion at each call,
and loops are not unrolled into hardware operations.
Operand widths are estimated from declarations and local expressions
without full Verilog elaboration.  Thus, $Q$ measures source-level compactness.

The TT-100 evaluator counts operators nested inside concatenations.
The PA-30 parser does not fully account for such nested operators.
The recovered PA-30 implementations contain neither bitwise operators
nor concatenations. Hence, these particular counting limitations do not
affect their reported costs.

Rewriting compares RTL candidates using the same evaluator and syntax-cost weights.
A lower $Q$ represents a more compact RTL design under the specified
syntax-cost weights. It does not guarantee lower synthesized area or delay.

\begin{table}[t]
\centering
\caption{Syntax-Cost Weights for PA-30 and TT-100}
\label{tab:cost_profiles}
\footnotesize
\setlength{\tabcolsep}{3pt}
\begin{tabular*}{0.98\columnwidth}{@{\extracolsep{\fill}}>{\raggedright\arraybackslash}p{0.72\columnwidth}rr@{}}
\toprule
Charged construct & PA-30 & TT-100 \\
\midrule
Binary arithmetic: ``+'', ``-'', ``*'', ``/'', ``**'', ``\%'' & 1 & 1 \\
Unary negation & 1 & 0 \\
Concatenation or replication, per symbol & 1 & 1 \\
Bit-select or part-select & 1 & 1 \\
Relational and equality operators & 1 & 1 \\
Logical NOT, AND, or OR & 1 & 1 \\
Bitwise NOT, AND, OR, XOR, or XNOR, per bit & 1 & 3 \\
Reduction AND, NAND, OR, NOR, XOR, or XNOR & 1 & 2 \\
Logical or arithmetic shift & 1 & 1 \\
Conditional expression ``?:'' or ``if'' statement & 1 & 1 \\
Nondefault ``case'', ``casex'', or ``casez'' item & 2 & 4 \\
Primitive-gate instance & 1 & 20 \\
\bottomrule
\end{tabular*}
\end{table}

\subsection{Cost-Guided Rewriting}\label{sec:refinement}
The LLM examines which expressions contribute to the syntax cost of verified
RTL and proposes expression
factorization, intermediate-width reduction, expression sharing, and
control-condition simplification. After RTL validity and equivalence checking, a rewritten
RTL candidate replaces the current implementation only if it passes both checks
and has a lower syntax cost. Unsuccessful attempts do not change the current
verified RTL candidate. At the end of a recovery run, RTLREX reports the
lowest-cost verified RTL candidate found, including when the run reaches its
time limit. The recovered RTL also
provides a starting point for arithmetic-aware resynthesis, whose
standard-cell area and delay are evaluated in \Cref{sec:cell_mapping}.

\section{Experimental Results}\label{sec:exp}

In this section, we describe the experimental setup and present recovery
results and comparisons with prior methods and the LLM agent baseline.
We then evaluate equivalence
verification, resynthesis, technology mapping, and generation time.

\subsection{Experimental Setup}\label{sec:setup}

\minisection{Benchmarks and Baselines}
We evaluate RTLREX on two benchmark suites. PA-30 comprises the 30
combinational primitive-gate netlists from Problem A of the
ICCAD 2022 CAD Contest~\cite{chou2022cadcontest,cadcontest2022problemA}.
TT-100 comprises the 100 complete truth tables \textit{ex200}--\textit{ex299}
from the official IWLS 2026 Contest, with 8--16 input bits and
4--32 output bits~\cite{iwls2026contest}. The input files match the public
repository revision recorded in the artifact.

For PA-30, we compare RTLREX with Hybrid-DATE~\cite{tang2024hybrid} and
WolFEx~\cite{ho2023wolfex}, using their published recovery coverage and
syntax costs on commonly solved benchmarks. We use WolFEx's published results
for its default (\texttt{-d}) configuration, which tries all extraction
techniques before expression optimization.

For TT-100, we compare RTLREX with an LLM agent baseline using the same
model, task requirements, generic tools, and independent verifier.
Only RTLREX has access to the domain-specific recovery
procedures and tools provided in the Agent Skill package. These procedures are
inspired by Hybrid-DATE~\cite{tang2024hybrid} and incorporate practical
human expertise in circuit reverse engineering. The package remains
unchanged throughout the experiment and contains no hard-coded benchmark
answers or benchmark-specific dispatch based on benchmark names or identifiers.

\minisection{Execution Settings}
\label{sec:matched_tt_protocol}\label{sec:updated_skill_protocol}
Each TT-100 benchmark is evaluated with RTLREX and the LLM agent baseline using
GPT-5.6 Sol with reasoning effort set to \texttt{xhigh}, a fresh context
for each run, and a 600-s generation time limit. Reported generation time includes
model interaction and tool calls, but excludes independent final validation.
RTLREX and the LLM agent baseline run on the same host machine with an Intel Core
i5-14400 CPU (10 cores), 128~GB of RAM, and Ubuntu 22.04.5 LTS.
Software versions and parallel execution settings are recorded in the
reproducibility materials.

The task prompts for both configurations specify the same recovery requirements
and bit-ordering convention for the IWLS truth-table files. Both configurations
also have access to Python, shell tools, Yosys, ABC, Icarus
Verilog, and Z3, together with a syntax-cost calculator and the same
verification tool. Note that this verification tool checks RTL validity and
equivalence but does not identify functions or generate RTL.
The prompts permit shared bit-level Boolean
implementations while discouraging input-enumerating lookup tables to
achieve high-level RTL recovery.

Within the same 600-s generation time limit, RTLREX allows at most two refinement
proposals within 120~s after the initial verified RTL candidate is produced.
Each proposal must pass independent validation and reduce syntax cost.
Repeated tool calls do not increase the proposal limit or extend the
refinement window. The implementation and task instructions
are described in \hyperref[app:prompts]{Appendix}.

\minisection{Independent Validation}
After each TT-100 run, we independently check the final RTL file for
validity and equivalence, and compute its syntax cost $Q$.
Whether a run finishes before the 600-s generation time limit or is terminated
at the limit, any final RTL file it has produced undergoes the same validity,
equivalence, and syntax-cost evaluation.
Yosys parses the Verilog source and resolves module references and parameter
values. It then checks the resulting circuit for structural errors, including
conflicting drivers, undriven signals, and combinational loops.
We use the command sequence
\texttt{read\_verilog}, \texttt{hierarchy -check -top}, \texttt{proc},
\texttt{opt}, and \texttt{check -assert}, with the configured top module.
These checks establish RTL validity, while equivalence verification
establishes functional correctness.

RTLREX and the LLM agent baseline use the same final-validation limits: 120~s for
Yosys RTL validity checking, 120~s for Icarus compilation, and 180~s for
complete-domain Verilog simulation. The bit-parallel evaluator processes
supported operations in Yosys's internal representation. Unsupported
operations, such as multiplication, are checked through complete-domain
simulation with Icarus Verilog. Acceptance requires matching values
on every output bit throughout the complete input domain.

\minisection{Evaluation Metrics}
\label{sec:metrics}
We report the number of validated RTL candidates ($V$) and the number of
high-level recoveries ($H$) using the criteria in \Cref{tab:endpoints}. All
benchmarks remain in the coverage denominator. Missing RTL candidates, detected
mismatches, and inconclusive verification results are recorded separately.
Any RTL candidate passing exact-equivalence verification and RTL validity
checking qualifies for $V$, including lookup-table RTL, bit-level Boolean RTL,
or a direct input-to-output connection when the circuit implements an
identity function.

\begin{table}[t]
\centering
\caption{RTL Recovery Criteria}
\label{tab:endpoints}
\footnotesize
\begin{tabular}{@{}l>{\raggedright\arraybackslash}p{0.82\columnwidth}@{}}
\toprule
Criterion & Requirements \\
\midrule
$V$ & Exact equivalence and successful RTL validity checking. \\
$H$ & All requirements of $V$ and an audited word- or field-level representation. PA-30 additionally requires at least 70\% syntax-cost reduction. \\
\bottomrule
\end{tabular}
\end{table}

A validated RTL candidate counts toward $H$ when source review confirms that
its RTL expresses word- or field-level computation. Primitive-gate networks,
copies of the original netlist, input-enumerating lookup tables, and
expanded bit-level Boolean implementations are excluded from $H$.
Conditional expressions are allowed within word-level RTL. For TT-100,
an implementation whose main computation is word-level is excluded from $H$
if it adds truth-table-derived corrections for explicitly enumerated input
values. These representation restrictions apply only to $H$.
RTL candidates that pass RTL validity checking and exact-equivalence verification
still count toward $V$ and remain eligible for syntax-cost comparison.

For PA-30, $H$ additionally requires at least 70\% syntax-cost reduction
relative to the original primitive-gate count, following the CAD Contest
specification~\cite{cadcontest2022problemA}. Each original primitive gate
has unit cost. We additionally review whether the RTL expresses word- or
field-level computation because the CAD Contest also permits primitive-gate
implementations. A reduction
in syntax cost alone does not establish high-level recovery.
For TT-100, $H$ requires the same source review of word- or field-level
computation, without a syntax-cost reduction threshold.

All final RTL candidates within each benchmark suite are scored using the same
evaluator and syntax-cost weights defined in \Cref{sec:syntax_cost}; costs are
compared separately within each suite. Since the published Hybrid-DATE
and WolFEx results do not cover all PA-30 benchmarks, each mean syntax-cost
comparison uses only the benchmarks recovered by RTLREX and the baseline
being compared. This yields \OursDateCommon{} benchmarks for Hybrid-DATE
and \OursWolfexCommon{} for WolFEx. For TT-100, we compare RTLREX with the
LLM agent baseline on benchmarks for which both methods satisfy $V$, and
separately on those for which both methods satisfy $H$.

\newcommand{\TTRecoveredFunctionGroupRows}{%
BF16 unary functions & \textit{ex200}--\textit{ex219} & 20 \\
FP16 unary functions & \textit{ex220}--\textit{ex239} & 20 \\
FP8 E4M3FN binary functions & \textit{ex240}--\textit{ex244} & 5 \\
FP8 E5M2 binary functions & \textit{ex245}--\textit{ex249} & 5 \\
FP4 dot product and fused multiply-add & \textit{ex250}, \textit{ex254} & 2 \\
Fixed-point dot products and sum & \textit{ex251}--\textit{ex253} & 3 \\
Signed and unsigned multiplication & \textit{ex255}--\textit{ex264} & 10 \\
Unsigned division & \textit{ex265}--\textit{ex269} & 5 \\
Unsigned square & \textit{ex270}--\textit{ex274} & 5 \\
Integer square root & \textit{ex275}--\textit{ex279} & 5 \\
Bit-level Boolean RTL & \textit{ex280}--\textit{ex299} & 20 \\
\midrule
\textbf{Total} & & \textbf{100} \\
}

\begin{table}[t]
\centering
\caption{Recovered Function Categories on TT-100}
\label{tab:tt100_recovered_functions}
\footnotesize
\setlength{\tabcolsep}{2pt}
\begin{tabular*}{0.98\columnwidth}{@{\extracolsep{\fill}}llr@{}}
\toprule
Recovered computation & Benchmarks & Count \\
\midrule
\TTRecoveredFunctionGroupRows
\bottomrule
\end{tabular*}
\end{table}

\newcommand{\MainResultsRows}{%
\rowcolor{RECOVERblue!7} PA-30 & \textbf{RTLREX} & 30/30 & 30/30 \\
\addlinespace[2pt]
\rowcolor{RECOVERblue!7} TT-100 & \textbf{RTLREX} & 100/100 & 80/100 \\
}
\newcommand{\SkillAblationRows}{%
\rowcolor{RECOVERblue!7} PA-30 & \textbf{RTLREX} & 30/30 & 30/30 \\
PA-30 & LLM agent baseline & 21/30 & 11/30 \\
\addlinespace[2pt]
TT-100 & \textbf{RTLREX} & 73/100 & 73/100 \\
TT-100 & LLM agent baseline & 0/100 & 0/100 \\
}

\setlength{\dbltextfloatsep}{20pt plus 2pt minus 2pt}
\begin{table*}[!t]
\centering
\begin{threeparttable}
\caption{Recovery Coverage, Operator Counts, and Syntax Cost on PA-30.\\
Bold marks the lowest cost, including ties, and RTLREX's complete recovery
coverage}
\label{tab:iccad_all_cases}
\footnotesize
\setlength{\tabcolsep}{7pt}
\renewcommand{\arraystretch}{1.14}
\begin{tabular*}{0.80\textwidth}{@{\extracolsep{\fill}}lrrrrrrr@{}}
\toprule
\multirow{2}{*}{Benchmark} & Original netlist & Hybrid-DATE~\cite{tang2024hybrid} &
WolFEx~\cite{ho2023wolfex} &
\multicolumn{4}{c}{\textbf{RTLREX}} \\
\cmidrule(lr){2-2}\cmidrule(lr){3-3}\cmidrule(lr){4-4}\cmidrule(l){5-8}
& \#Gates ($= Q$) & Cost ($Q$) & Cost ($Q$) & Add & Mul & Cond. & Cost ($Q$) \\
\midrule
\RecoverCaseRows
\bottomrule
\end{tabular*}
\end{threeparttable}
\end{table*}

\subsection{Recovery Coverage and RTL Representation}
\label{sec:recover_endpoint_profile}
\label{sec:exactness_results}\label{sec:representation_results}

RTLREX produces an RTL candidate for every benchmark in PA-30 and TT-100.
All PA-30 benchmarks satisfy the validated-RTL criterion $V$ and the
high-level-RTL criterion $H$. On TT-100, all benchmarks satisfy $V$, and 80
satisfy $H$. The remaining 20 benchmarks use bit-level Boolean RTL,
satisfying $V$ but not exposing recovered word-level structure.

The 80 high-level TT-100 RTL candidates expose word- or field-level computation,
comprising 52 floating-point, 25 integer, and three fixed-point functions.
\Cref{tab:tt100_recovered_functions} lists the recovered computations
and corresponding TT-100 benchmarks. The categories in
\Cref{tab:tt100_recovered_functions} describe functions identified by RTLREX
rather than labels supplied with the benchmarks.

FP16, FP8, and FP4 denote 16-, 8-, and 4-bit floating-point formats,
respectively. In the FP8 labels, the numbers following E and M specify the
exponent and fraction widths, respectively, and FN indicates a format that
supports NaNs but not infinities.

For all PA-30 benchmarks, formal proofs establish that the
recovered RTL and the original netlist agree on every corresponding output
bit for every input combination. For TT-100, complete-domain RTL simulation establishes
exact equivalence for all benchmarks, checking a total of 4\,548\,864 input
vectors across the suite, including the 20 bit-level Boolean implementations.
All recovered RTL also passes Yosys RTL validity checking. Verification methods
and proof results are detailed in \Cref{sec:verification_analysis}.

Per-benchmark recovered functions, interface widths,
$H$ classifications, and syntax costs are available in the software companion
described in \hyperref[app:prompts]{Appendix}.
The $H$ criteria assess whether RTL expresses recovered word- or field-level
computation. The RTL designs for \textit{test19}, \textit{test24}, and
\textit{ex207} satisfy this requirement, whereas bit-level Boolean
implementations do not.

\subsection{Comparison with Prior Methods}\label{sec:baseline_comparison}

\Cref{tab:iccad_all_cases} compares RTLREX with Hybrid-DATE and WolFEx
on PA-30 using the contest syntax-cost weights specified in \Cref{sec:syntax_cost}.

Hybrid-DATE reports successful recovery for 23 benchmarks, and WolFEx for 26
benchmarks. RTLREX recovers and verifies all \TableFiveCertified{} benchmarks. Every
recovered implementation reduces syntax cost by at least 70\% relative to
the original gate count, meeting the contest requirement.
In \Cref{tab:iccad_all_cases}, Add and Mul represent the numbers of ``+''
and ``*'' operators in the recovered RTL, respectively. Cond. represents
the total number of ``if'' statements and ``?:'' conditional expressions. The total syntax cost
$Q$ includes all charged constructs. In the published baseline results,
Failed and Timeout denote unsuccessful recovery and exceeding the time
limit, respectively.

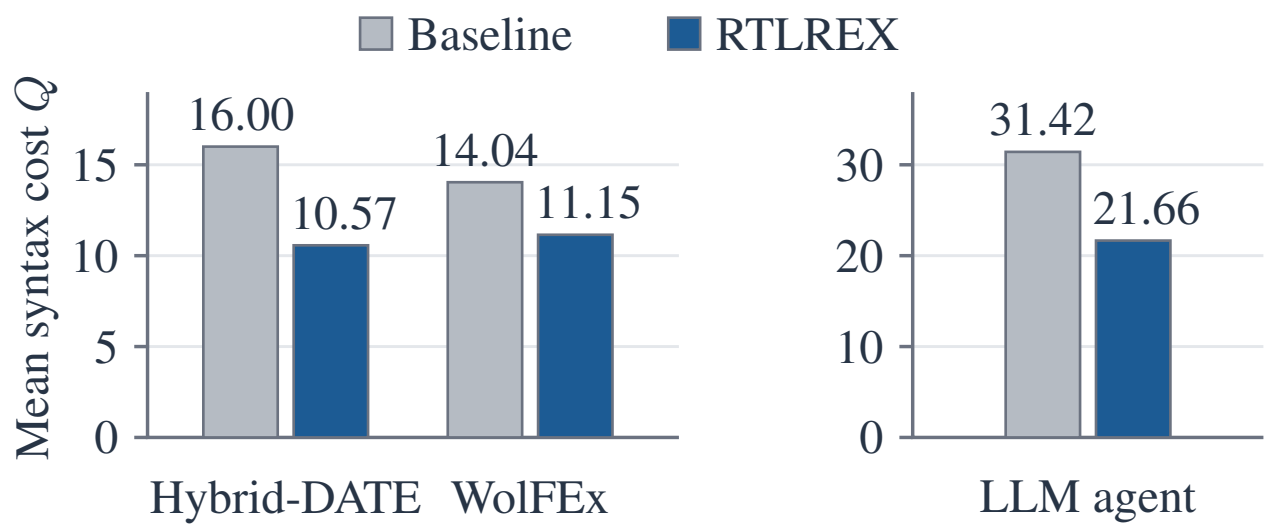
\begin{figure}[t]
\centering
\definecolor{costink}{HTML}{273445}
\definecolor{costaxis}{HTML}{6B7280}
\definecolor{costgrid}{HTML}{E4E7EB}
\definecolor{costbaseline}{HTML}{B5BBC3}
\definecolor{costrtlrex}{HTML}{1C5B94}
\tikzset{recovery cost plot/.style={x=1bp,y=1bp,
  font=\normalfont\rmfamily\fontsize{9bp}{10.8bp}\selectfont,
  text=costink,inner sep=0pt,outer sep=0pt}}
\newcommand{\RecoveryCostAxes}[4]{%
  \begin{scope}[shift={(#1,43)}]
    \foreach \tick in {#4} {
      \pgfmathsetmacro{\tickheight}{65*\tick/#3}
      \draw[costgrid,line width=.55bp] (0,\tickheight) -- (#2,\tickheight);
      \draw[costaxis,line width=.65bp] (-2.5,\tickheight) -- (0,\tickheight);
      \node[anchor=east] at (-5.5,\tickheight) {\tick};
    }
    \draw[costaxis,line width=.65bp] (0,65) -- (0,0) -- (#2,0);
  \end{scope}}
\newcommand{\RecoveryCostBar}[5]{%
  \pgfmathsetmacro{\barheight}{65*#2/#3}
  \draw[fill=#4,draw=costaxis,line width=.5bp]
    ({#1-7},43) rectangle ({#1+7},{43+\barheight});
  \node[anchor=south] at ({#1+#5},{46+\barheight}) {#2};
}
\newcommand{\RecoveryLegend}{%
  \begin{tikzpicture}[recovery cost plot]
    \path[use as bounding box] (0,0) rectangle (104,10);
    \draw[fill=costbaseline,draw=costaxis,line width=.5bp] (0,2) rectangle (6,8);
    \node[anchor=west] at (9,5) {Baseline};
    \draw[fill=costrtlrex,draw=costaxis,line width=.5bp] (58,2) rectangle (64,8);
    \node[anchor=west] at (67,5) {RTLREX};
  \end{tikzpicture}%
}
\newcommand{\RecoveryTTPanel}{%
  \begin{tikzpicture}[recovery cost plot]
    \path[use as bounding box] (0,22) rectangle (90,112);
    \begin{pgfinterruptboundingbox}
      \RecoveryCostAxes{18}{66}{38}{0,10,20,30}
      \RecoveryCostBar{42.5}{\UpdatedCostHNoSkillMean}{38}{costbaseline}{0}
      \RecoveryCostBar{59.5}{\UpdatedCostHWithSkillMean}{38}{costrtlrex}{2.5}
      \node[anchor=north] at (51,35) {LLM agent};
    \end{pgfinterruptboundingbox}
  \end{tikzpicture}%
}
\newcommand{\RecoveryPAPanel}{%
  \begin{tikzpicture}[recovery cost plot]
    \path[use as bounding box] (0,22) rectangle (144,112);
    \begin{pgfinterruptboundingbox}
      \RecoveryCostAxes{26}{100}{19}{0,5,10,15}
      \RecoveryCostBar{43.5}{\OursDateRhsMean}{19}{costbaseline}{0}
      \RecoveryCostBar{60.5}{\OursDateLhsMean}{19}{costrtlrex}{2.5}
      \RecoveryCostBar{89.5}{\OursWolfexRhsMean}{19}{costbaseline}{0}
      \RecoveryCostBar{106.5}{\OursWolfexLhsMean}{19}{costrtlrex}{2.5}
      \node[anchor=north] at (52,35) {Hybrid-DATE};
      \node[anchor=north] at (98,35) {WolFEx};
      \node[rotate=90] at (5.2,75) {Mean syntax cost $Q$};
    \end{pgfinterruptboundingbox}
  \end{tikzpicture}%
}
\newcommand{\RecoveryPanels}{%
  \begin{minipage}{252bp}
    \centering
    \RecoveryLegend\par\vspace{2bp}
    \hspace*{8bp}%
    \subfloat[PA-30\label{fig:recovery_pa}]{\RecoveryPAPanel}%
    \hspace{8bp}%
    \subfloat[TT-100\label{fig:recovery_tt}]{\RecoveryTTPanel}%
    \hspace*{2bp}
  \end{minipage}%
}
\resizebox{\columnwidth}{!}{\RecoveryPanels}
\captionsetup{justification=justified}
\caption{Mean syntax cost on common successful benchmarks.
(a) PA-30: paired comparisons with Hybrid-DATE on \OursDateCommon{} common
benchmarks and WolFEx on \OursWolfexCommon{} common benchmarks.
(b) TT-100: \UpdatedCostHCommon{} common high-level recoveries.}
\label{fig:recovery_summary}
\end{figure}

On the \OursDateCommon{} benchmarks recovered by both RTLREX and Hybrid-DATE,
RTLREX reduces mean syntax cost from \OursDateRhsMean{} to
\OursDateLhsMean{}, a reduction of \OursDateMeanReduction\%,
as shown in \Cref{fig:recovery_summary}(a).
RTLREX achieves lower syntax cost on \OursDateWins{} benchmarks and equal
syntax cost on the remaining \OursDateTies{} benchmarks.

On the \OursWolfexCommon{} benchmarks recovered by both RTLREX and WolFEx,
RTLREX reduces mean syntax cost from \OursWolfexRhsMean{} to
\OursWolfexLhsMean{}, a reduction of \OursWolfexMeanReduction\%.
RTLREX achieves lower syntax cost on \OursWolfexWins{} benchmarks and equal
syntax cost on the remaining \OursWolfexTies{} benchmarks.

RTLREX also recovers an equivalent RTL implementation of \textit{test27} with a
syntax cost of 49, while Hybrid-DATE reports failure and WolFEx reports
a timeout.

\subsection{Comparison with the LLM Agent Baseline}\label{sec:skill_ablation}

We compare RTLREX with the LLM agent baseline on all TT-100 benchmarks. Both methods use GPT-5.6 Sol with reasoning effort
set to \texttt{xhigh} and a 600-s generation time limit per benchmark, and both receive the
explicit bit-ordering convention for the IWLS truth tables. All final RTL
implementations are evaluated using the uniform verification and scoring
procedure described in \Cref{sec:matched_tt_protocol}.

As shown in \Cref{tab:skill_ablation}, RTLREX produces validated RTL for all
TT-100 benchmarks, including high-level RTL for \UpdatedSkillH{} benchmarks.
The LLM agent baseline produces
validated RTL for \RetainedNoSkillV{} benchmarks, including \RetainedNoSkillH{}
with high-level RTL descriptions. RTLREX therefore yields
19 additional benchmarks with validated RTL and six additional benchmarks
with high-level RTL.
For the remaining 20 benchmarks, RTLREX generates bit-level Boolean RTL
that satisfies $V$ but not $H$.

\begin{table}[t]
\centering
\caption{Recovery Coverage on TT-100}
\label{tab:skill_ablation}
\label{tab:updated_skill_retained}
\footnotesize
\setlength{\tabcolsep}{3pt}
\begin{tabular*}{0.90\columnwidth}{@{\extracolsep{\fill}}lccc@{}}
\toprule
Configuration & Trials & $V$ & $H$ \\
\midrule
\UpdatedSkillRetainedCoverageRows
\bottomrule
\end{tabular*}
\end{table}

\begin{table}[t]
\centering
\caption{Syntax-Cost Comparison on Common TT-100 Benchmarks.\\
Wins and losses indicate lower and higher RTLREX syntax costs than the LLM agent baseline, respectively}
\label{tab:matched_tt_cost}
\label{tab:updated_skill_retained_cost}
\footnotesize
\setlength{\tabcolsep}{3pt}
\begin{tabular*}{\columnwidth}{@{\extracolsep{\fill}}lccccc@{}}
\toprule
\multirow{2}{*}{Set} & \multirow{2}{*}{Common} &
\multicolumn{3}{c}{RTLREX vs. baseline} &
\multirow{2}{*}{\shortstack{Mean cost\\reduction}} \\
\cmidrule(lr){3-5}
& & Wins & Ties & Losses & \\
\midrule
\UpdatedSkillRetainedCostRows
\bottomrule
\end{tabular*}
\end{table}

Three LLM agent baseline implementations, \textit{ex223}, \textit{ex227}, and
\textit{ex228}, express the main computation using high-level arithmetic RTL,
but override its result for a small set of explicitly enumerated input
values. In \textit{ex223}, one input value selects a stored output constant,
forming a one-entry lookup. In \textit{ex227}, the RTL substitutes predefined
output values for three explicitly listed positive inputs and their negative
counterparts, covering six input values in total. In \textit{ex228}, the RTL
handles one specified positive input and its negative counterpart by decreasing
the computed 15-bit exponent-and-fraction encoding by one.
These input-specific overrides differ from general numerical-format rules for
zero, NaN, or infinity. The RTL implementations for these three benchmarks pass
complete-domain equivalence verification and RTL validity checking.
They therefore count toward $V$ and are included in the syntax-cost comparison
on common validated benchmarks. Although their main
computation is high-level, the input-specific overrides exclude them
from $H$.

\Cref{tab:skill_ablation} reports each method's recovery coverage, whereas
\Cref{tab:matched_tt_cost} compares syntax cost only on benchmarks satisfying
the same criterion for both methods. The baseline's \RetainedNoSkillV{}
validated recoveries and \RetainedNoSkillH{} high-level recoveries are all also
achieved by RTLREX, yielding common sets of \UpdatedCostVCommon{} and
\UpdatedCostHCommon{} benchmarks, respectively. All final RTL implementations are scored
using the same syntax-cost evaluator and weights. Mean cost reduction is
calculated from the two mean costs on the same set of benchmarks.
Wins, Ties, and Losses count benchmarks on which RTLREX has lower, equal,
or higher syntax cost than the LLM agent baseline, respectively.

On the \UpdatedCostVCommon{} common validated benchmarks, RTLREX reduces
mean syntax cost from \UpdatedCostVNoSkillMean{} to
\UpdatedCostVWithSkillMean{}, a reduction of \UpdatedCostVMeanReduction\%.
This comparison includes four bit-level Boolean implementations from the
LLM agent baseline and the three implementations with input-specific
corrections. The only higher-cost benchmark is \textit{ex228}, with costs of 61
for RTLREX and 57 for the LLM agent baseline. On the \UpdatedCostHCommon{} common
benchmarks with high-level RTL, mean cost falls from \UpdatedCostHNoSkillMean{} to
\UpdatedCostHWithSkillMean{}, a reduction of \UpdatedCostHMeanReduction\%,
as shown in \Cref{fig:recovery_summary}(b).
Every benchmark in this set has lower or equal cost with RTLREX.

\KeepWithNext{4\baselineskip}
\subsection{Equivalence Verification}\label{sec:verification_analysis}

For PA-30, exact equivalence is established through CEC, bit-vector SMT, or
arithmetic proofs, as described in \Cref{sec:verification}.
These proofs establish that the recovered RTL and original netlist agree on
every corresponding output bit for all input combinations, without exhaustively
simulating the complete input domain. The common verification procedure uses
ABC's \texttt{\&cec}, \texttt{dcec}, and \texttt{cec} engines and treats
contradictory equivalence and inequivalence results as inconclusive.
Its limits are 60~s per ABC engine, 120~s per Yosys invocation, and 45~s for
the algebraic fallback, with a maximum of 100\,000 polynomial terms.
For benchmarks not resolved within these limits, additional proof procedures
use time and resource limits specified for each proof.

\newcommand{\ProofDirectCEC}{15}
\newcommand{\ProofArithmetic}{6}
\newcommand{\ProofCombined}{9}
\newcommand{\ProofSMT}{1}
\newcommand{\ProofMethodTotal}{31}
\newcommand{\ProofRouteRows}{%
Direct ABC CEC & 15 \\
Arithmetic replay & 6 \\
Combined arithmetic/Boolean plans & 9 \\
Bit-vector SMT & 1 \\
}

\Cref{tab:proof_routes} reports complete-interface proof methods for the
recovered RTL across all PA-30 benchmarks, comprising \ProofDirectCEC{} direct
CEC proofs, \ProofArithmetic{} arithmetic proofs, and \ProofCombined{}
combined proofs. Bit-vector SMT also independently verifies \textit{test23},
which has a CEC proof. The counts therefore overlap and sum to
\ProofMethodTotal{} across 30 distinct benchmarks.
Each benchmark is counted once per proof method. Combined proofs use
independently checked arithmetic relations, CEC or reduced ordered binary
decision diagram (ROBDD) checks, and complete coverage of selected control-bit
combinations as needed. Proved intermediate relations reconnect
these obligations to the complete original interfaces.

\begin{table}[t]
\centering
\caption{Equivalence Proof Methods Used on PA-30}
\label{tab:proof_routes}
\footnotesize
\begin{tabular*}{0.86\columnwidth}{@{\extracolsep{\fill}}lr@{}}
\toprule
Proof method & Count \\
\midrule
\ProofRouteRows
\midrule
Total (including overlap) & \ProofMethodTotal{} \\
\bottomrule
\end{tabular*}
\end{table}

The common verification procedure certifies 18 RTL candidates, comprising 15 CEC
proofs and three arithmetic proofs. The remaining 12 RTL candidates are verified
using additional arithmetic or combined proof procedures with separately
specified limits. For example, \textit{test04} uses a one-million-term cap for
arithmetic reduction ordered by output-bit significance. For \textit{test26},
checking every assignment of selected control bits establishes internal
arithmetic relations, and arithmetic and ROBDD checks verify the remaining
output obligations. The \textit{test22} example
in \hyperref[sec:netlist_proof_example]{Section~\ref*{sec:verification}}
illustrates how a verified intermediate circuit
connects the original netlist to the recovered RTL. TT-100 uses complete-domain
simulation throughout.

\KeepWithNext{5\baselineskip}
\subsection{Resynthesis and Technology Mapping}\label{sec:cell_mapping}

Following the motivation in \Cref{fig:motivation}, we examine whether
recovered high-level RTL provides useful alternative implementations for
resynthesis. We evaluate one final recovered implementation for each
PA-30 benchmark against direct mapping of the original gate-level netlist.

Recovering word-level operations exposes arithmetic information that can be
used before the computation is lowered into Boolean logic. In this
experiment, uniform transformations exploit symmetric partial products in
unsigned squares, implement constant multiplications as signed-digit
shift-and-add networks, and lower word-level comparisons into grouped
lookahead logic. Booth encoding is enabled during synthesis. These
transformations use the recovered operations and operand widths to construct
the logic presented to subsequent optimization and technology mapping,
without benchmark-specific dispatch.

Both sides use the same Nangate45 typical-corner
library~\cite{nangate45library} and ABC mapping constraints: a 2~ns delay
target, a BUF\_X1 driving cell, and a 10~fF output load.
OpenSTA~\cite{opensta} evaluates pre-layout timing with ideal interconnect,
the same driving cell and load, a 0.05~ns driver-input transition,
a 10~ns virtual clock, and zero external input and output delays.
Standard-cell area and input-to-output delay are evaluated separately from
syntax cost $Q$.

\newcommand{\StandardCellRows}{%
\textit{test01} & 32.5 & \textbf{31.7} & \textbf{2.46} & 0.242 & 0.274 & $-13.22$ \\
\textit{test02} & 45.2 & \textbf{43.6} & \textbf{3.53} & 0.311 & 0.395 & $-26.94$ \\
\textit{test03} & 273.4 & \textbf{260.4} & \textbf{4.77} & 0.946 & \textbf{0.877} & \textbf{7.32} \\
\textit{test04} & 4199.9 & \textbf{2453.6} & \textbf{41.58} & 2.007 & \textbf{1.991} & \textbf{0.84} \\
\textit{test05} & 1102.3 & 1245.7 & $-13.01$ & 1.460 & 1.609 & $-10.26$ \\
\textit{test06} & 3512.8 & 3942.9 & $-12.24$ & 3.017 & \textbf{2.452} & \textbf{18.71} \\
\textit{test07} & 1357.7 & 1496.8 & $-10.25$ & 1.904 & \textbf{1.814} & \textbf{4.74} \\
\textit{test08} & 1521.3 & 1536.2 & $-0.98$ & 2.786 & \textbf{2.125} & \textbf{23.72} \\
\textit{test09} & 3953.3 & 4556.6 & $-15.26$ & 2.093 & \textbf{2.031} & \textbf{2.95} \\
\textit{test10} & 1505.3 & 1637.5 & $-8.78$ & 1.889 & 1.915 & $-1.35$ \\
\textit{test11} & 1172.5 & 1332.9 & $-13.68$ & 1.659 & 1.673 & $-0.82$ \\
\textit{test12} & 9505.2 & 9530.5 & $-0.27$ & 1.921 & 1.987 & $-3.46$ \\
\textit{test13} & 461.5 & \textbf{257.8} & \textbf{44.15} & 0.776 & 0.778 & $-0.27$ \\
\textit{test14} & 32.7 & 39.1 & $-19.51$ & 0.232 & 0.302 & $-30.11$ \\
\textit{test15} & 2378.8 & 2526.5 & $-6.21$ & 1.983 & \textbf{1.976} & \textbf{0.35} \\
\textit{test16} & 342.6 & 377.2 & $-10.09$ & 1.533 & \textbf{1.488} & \textbf{2.95} \\
\textit{test17} & 532.0 & \textbf{524.6} & \textbf{1.40} & 1.788 & \textbf{1.605} & \textbf{10.23} \\
\textit{test18} & 62.8 & \textbf{58.5} & \textbf{6.78} & 0.371 & \textbf{0.354} & \textbf{4.47} \\
\textit{test19} & 1173.3 & \textbf{671.1} & \textbf{42.80} & 0.645 & \textbf{0.525} & \textbf{18.62} \\
\textit{test20} & 3487.8 & \textbf{3434.6} & \textbf{1.53} & 1.431 & 1.615 & $-12.87$ \\
\textit{test21} & 1344.9 & \textbf{1090.1} & \textbf{18.95} & 1.260 & 1.273 & $-1.08$ \\
\textit{test22} & TLE & \textbf{70883.9} & -- & TLE & \textbf{3.578} & -- \\
\textit{test23} & 249.5 & 255.4 & $-2.35$ & 1.338 & \textbf{1.042} & \textbf{22.12} \\
\textit{test24} & 12344.5 & \textbf{12003.5} & \textbf{2.76} & 2.073 & 2.187 & $-5.52$ \\
\textit{test25} & 72.6 & 84.3 & $-16.12$ & 0.450 & \textbf{0.409} & \textbf{9.15} \\
\textit{test26} & 10631.5 & 10760.8 & $-1.22$ & 2.827 & 2.890 & $-2.22$ \\
\textit{test27} & 1809.6 & \textbf{1630.0} & \textbf{9.92} & 1.351 & \textbf{1.207} & \textbf{10.67} \\
\textit{test28} & 5729.4 & \textbf{5708.4} & \textbf{0.37} & 1.672 & \textbf{1.523} & \textbf{8.91} \\
\textit{test29} & 350.1 & 410.4 & $-17.25$ & 1.401 & \textbf{1.383} & \textbf{1.33} \\
\textit{test30} & 10228.2 & \textbf{10046.3} & \textbf{1.78} & 2.077 & 2.152 & $-3.61$ \\
\midrule
\textbf{Avg.} & 2738.4 & \textbf{2687.8} & \textbf{1.23} & 1.498 & \textbf{1.443} & \textbf{1.22} \\
}

\begin{table}[t]
\centering
\caption{Standard-Cell Area and Delay on PA-30.\\
Bold: better results. TLE: mapping timeout}
\label{tab:cell_mapping}
\footnotesize
\setlength{\tabcolsep}{1.5pt}
\begin{tabular*}{0.98\columnwidth}{@{\extracolsep{\fill}}l*{6}{r}@{}}
\toprule
\multirow{2}{*}{Benchmark} &
\multicolumn{3}{c}{Cell area ($\mu$m$^2$)} & \multicolumn{3}{c}{Delay (ns)} \\
\cmidrule(lr){2-4}\cmidrule(l){5-7}
& Original & RTLREX & $\downarrow$ (\%) &
Original & RTLREX & $\downarrow$ (\%) \\
\midrule
\StandardCellRows
\bottomrule
\end{tabular*}
\end{table}

For each benchmark in \Cref{tab:cell_mapping}, we report the percentage reduction
relative to the original netlist. The arithmetic mean of these per-benchmark
reductions over the 29 benchmarks with completed measurements on both sides
is 1.23\% for standard-cell area and 1.22\% for input-to-output delay.
The original netlist of \textit{test22} exceeds the 600-s mapping limit,
while its recovered RTL is successfully mapped.
These pre-layout area and delay measurements exclude placement and
extracted interconnect parasitics.

Individual circuits exhibit different area--delay tradeoffs.
For example, \textit{test04} reduces area by 41.6\% and delay by 0.8\%,
while \textit{test06} reduces delay by 18.7\% with a 12.2\% area increase.
These results show that recovered high-level RTL can provide useful
alternatives for resynthesis when combined with suitable arithmetic lowering.
The benefits depend on the circuit and synthesis flow. The measurements
evaluate the final recovered implementations under this arithmetic-aware
synthesis flow.

\subsection{Runtime and Failure Analysis}\label{sec:runtime_failures}

RTLREX produces RTL that passes independent RTL validity and complete-domain
equivalence checks for all TT-100 benchmarks, compared with
\RetainedNoSkillV{} benchmarks for the LLM agent baseline. For the baseline,
\RetainedNoSkillMissingFinal{} runs produce no RTL, one RTL candidate is disproved
by a counterexample, and one remains inconclusive at the verification time limit.
All TT-100 benchmarks remain in the coverage denominator.

Under the same 600-s generation time limit, RTLREX reduces the number of runs reaching
the limit from \RetainedNoSkillTimeouts{} to \UpdatedSkillTimeouts{}.
For \textit{ex287}, the generated RTL passes subsequent independent verification.

Across all TT-100 benchmarks, RTLREX reduces mean per-benchmark generation time from
\RetainedNoSkillGenerationMean{}~s to \UpdatedSkillGenerationMean{}~s,
a reduction of \UpdatedGenerationMeanReduction\%.
On the \UpdatedCostVCommon{} common validated benchmarks, mean generation time
falls by \UpdatedRuntimeVMeanReduction\%, from
\UpdatedRuntimeVNoSkillMean{}~s to \UpdatedRuntimeVWithSkillMean{}~s.
On the \UpdatedCostHCommon{} common benchmarks with high-level RTL, it falls from
\UpdatedRuntimeHNoSkillMean{}~s to \UpdatedRuntimeHWithSkillMean{}~s,
a reduction of \UpdatedRuntimeHMeanReduction\%.
These comparisons show faster recovery even when both methods produce
validated RTL.  Generation time includes model interaction and tool calls,
including trials that reach the time limit, and excludes independent final
validation.

Per-benchmark recovery outcomes, verification results, syntax costs, and generation
times are available in the companion repository.

\ifdim\baselinestretch pt<1.5pt\relax
  \KeepWithNext{16\baselineskip}
\fi
\section{Conclusion}\label{sec:conclusion}

RTLREX combines LLM-guided analysis and RTL revision with domain-specific
tools and independent verification to recover compact RTL from gate-level netlists or complete
truth tables. Recovered arithmetic expressions and verified intermediate
circuits support exact netlist equivalence proofs. The framework proves
equivalence for all PA-30 benchmarks and produces
validated RTL for all TT-100 benchmarks, including
\UpdatedSkillH{} high-level RTL descriptions. It achieves higher recovery
coverage than the evaluated baselines and lower mean syntax cost on the
benchmarks successfully recovered by the methods being compared.
The recovered high-level RTL supports rewriting,
resynthesis, and technology mapping for area and delay optimization.

\FloatBarrier
\KeepWithNext{8\baselineskip}
\appendix[Task Instructions and Software]\label[appendix]{app:prompts}

The MIT-licensed RTLREX software companion is available at
\url{https://github.com/wuqian-tang/RTLREX}.
It provides the \texttt{circuit-reverse-rtl} Agent Skill, recovery and rewriting
tools, equivalence checkers, tests, and synthetic examples, together with
installation instructions and dependencies.

For TT-100, the companion provides the task prompts, experiment configurations,
and Agent Skill source snapshot used in the reported evaluation.
Final evaluation follows \Cref{sec:matched_tt_protocol}.

The experiment packages include final RTL candidates, verification summaries,
cost configurations, tool versions, per-benchmark measurements, and scripts
that recompute the reported aggregates
without rerunning RTL recovery or equivalence verification.
The TT-100 package includes \texttt{recovered\_functions.csv}, which provides
per-benchmark recovered functions, interface widths, $V/H$ classifications, and
syntax costs. The BF16 exponential example in \path{examples/bf16-exp/} provides
the initial and rewritten RTL, verification results, syntax-cost measurements,
and commands for checking both implementations.
The repository includes the TT-100 truth tables with source attribution
and the original license~\cite{iwls2026contest}.
The repository also includes the original PA-30 gate-level netlists with
official source attribution and file checksums~\cite{cadcontest2022problemA}.
Third-party EDA tools must be obtained separately.
The repository provides a script for downloading the Nangate45 library
used for mapping.  Regenerating agent-based RTL recovery additionally
requires model access configured according to \Cref{sec:matched_tt_protocol}.

\FloatBarrier
\section*{Acknowledgment}

The authors used OpenAI Codex to assist with manuscript polishing and
consistency checks. The authors reviewed all content edited or generated
using artificial intelligence tools, verified all reported numerical
results, and take full responsibility for the manuscript.

{
\renewcommand{\IEEEbibitemsep}{0.85pt plus 0.5pt}
\bibliographystyle{IEEEtran}
\bibliography{ref/reference}
}
\begingroup
\makeatletter
\def\@IEEEBIOskipN{54pt}
\makeatother

\begin{IEEEbiography}[{\includegraphics[width=1in,height=1.25in,clip]{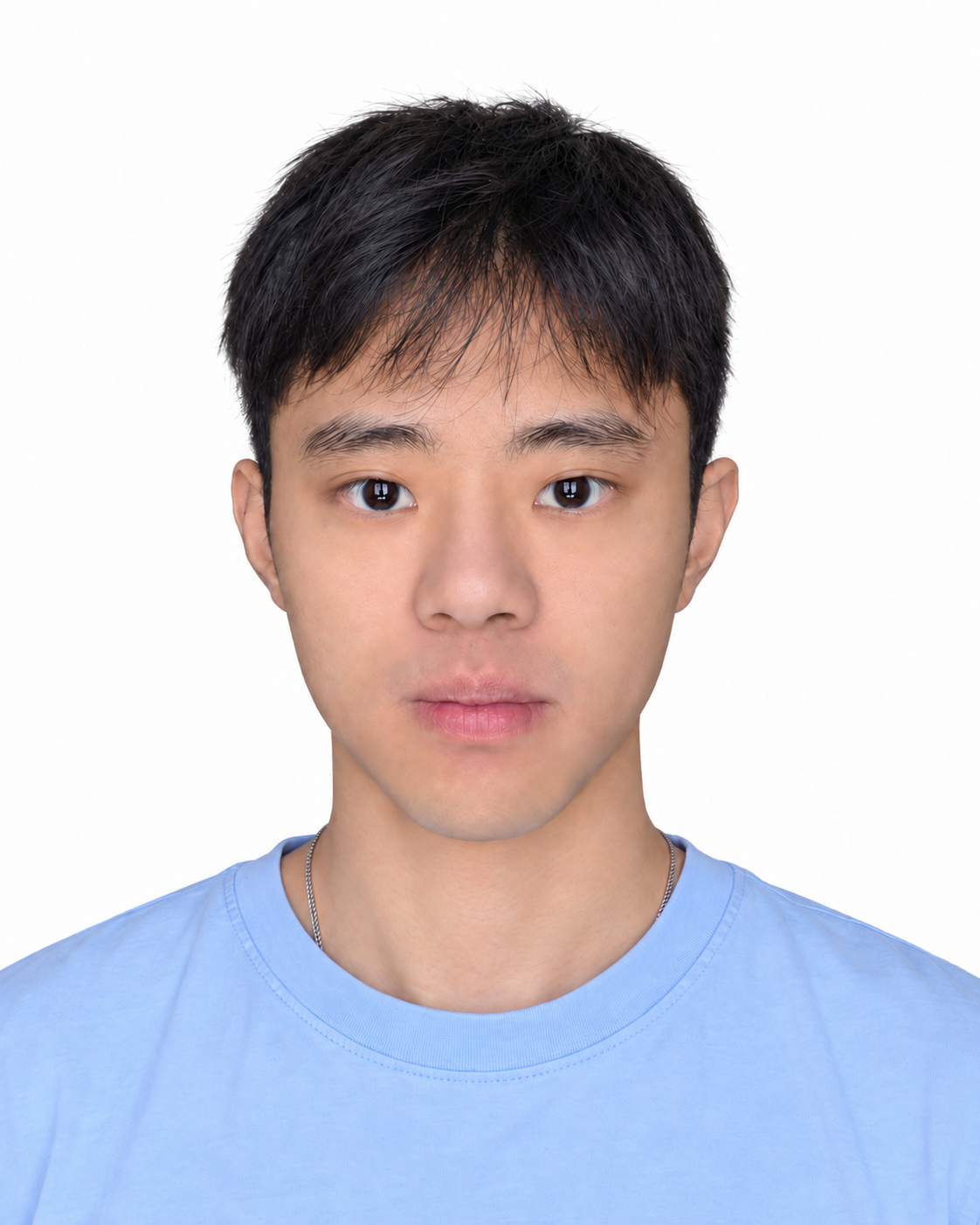}}]{Wuqian Tang}
(Student Member, IEEE) received the B.S. degree in computer science and
information engineering from Ming Chuan University, Taoyuan, Taiwan, in
2021, and the M.S. degree in computer science from National Tsing Hua
University, Hsinchu, Taiwan, in 2023, where he is currently pursuing the
Ph.D. degree.  His research interests include logic synthesis and machine
learning for electronic design automation.
He has received more than ten ICCAD/\allowbreak CADathlon/\allowbreak
MLCAD/\allowbreak IWLS/\allowbreak EDAthon contest awards, including
three first-place awards, and is a recipient of the
Synopsys - Prof. Chung Laung Liu Ph.D.
Scholarship.
\end{IEEEbiography}

\begin{IEEEbiography}[{\includegraphics[width=1in,height=1.25in,clip]{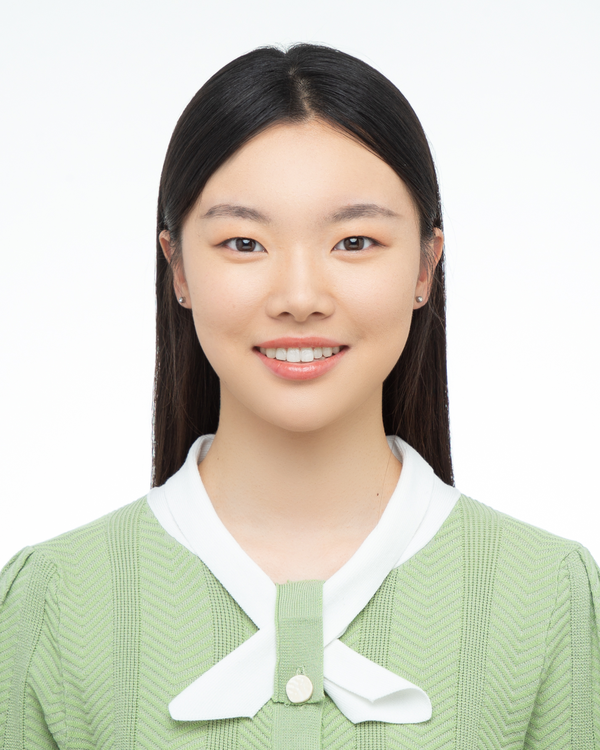}}]{Fangzhou Liu}
received the B.Eng. degree from the School of Electronic Science and
Engineering, Nanjing University, Nanjing, China, in 2023.  She is currently
pursuing the Ph.D. degree with the Department of Computer Science and
Engineering, The Chinese University of Hong Kong, Hong Kong.  Her research
interests include machine learning for electronic design automation and logic
synthesis optimization.
\end{IEEEbiography}

\ifdim\baselinestretch pt<1.5pt\relax
  \newpage
\fi

\begin{IEEEbiography}[{\includegraphics[width=1in,height=1.25in,clip]{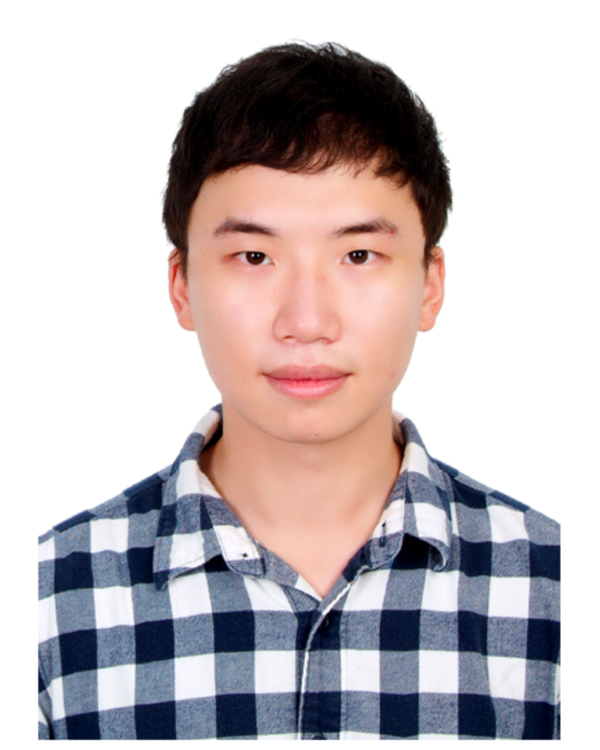}}]{Ruijie Wang}
received the B.S. degree from the Department of Applied Mathematics, Feng Chia
University, Taichung, Taiwan, in 2020.  He is currently pursuing the Ph.D.
degree with the Department of Computer Science, National Tsing Hua University,
Hsinchu, Taiwan.  His research interests include logic synthesis and hardware
security.
\end{IEEEbiography}

\begin{IEEEbiography}[{\includegraphics[width=1in,height=1.25in,clip]{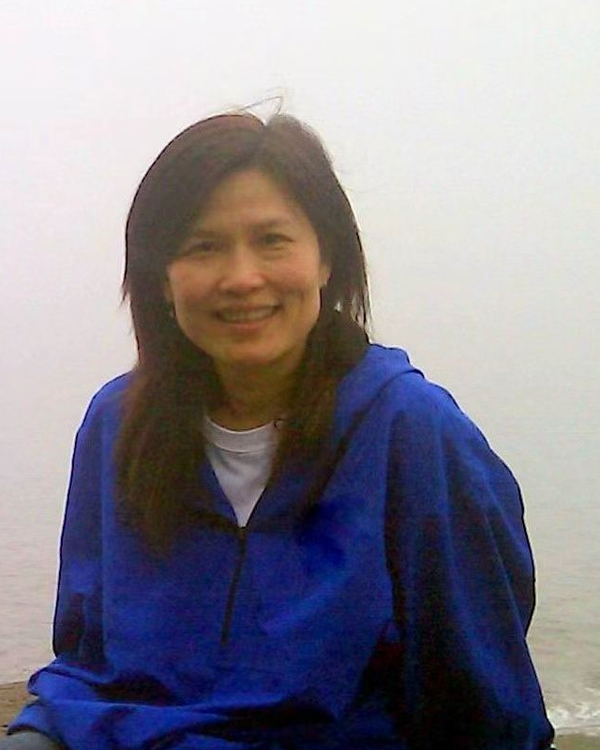}}]{TingTing Hwang}
(Member, IEEE) received the M.S. and Ph.D. degrees in computer science from Pennsylvania
State University, University Park, PA, USA, in 1986 and 1990, respectively.
She is currently a Distinguished Professor with the Department of Computer Science, National
Tsing Hua University, Hsinchu, Taiwan.  Her research interests include logic
synthesis and optimization and high-level synthesis.
\end{IEEEbiography}

\begin{IEEEbiography}[{\includegraphics[width=1in,height=1.25in,clip]{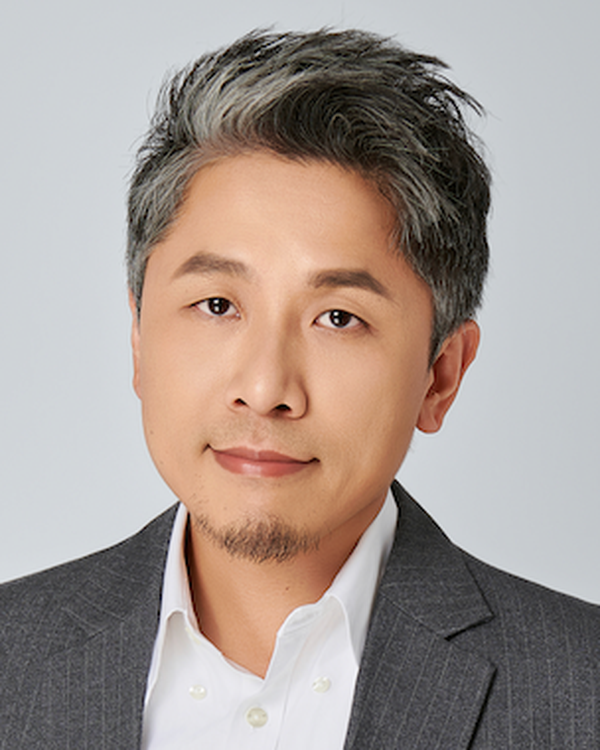}}]{Bei Yu}
(Senior Member, IEEE) received the Ph.D. degree from The University of Texas at
Austin, Austin, TX, USA, in 2014.  He is currently a Professor with the
Department of Computer Science and Engineering, The Chinese University of Hong
Kong, Hong Kong.  He has served as the Technical Program Committee Chair of
the ACM/IEEE Workshop on Machine Learning for CAD in 2019 and the International
Symposium of EDA in 2025, and on journal editorial boards and conference
committees.  He has received eleven Best Paper Awards, ten ICCAD/ISPD/IWLS
contest awards, the IEEE CEDA Ernest S. Kuh Early Career Award in 2021, the DAC
Under-40 Innovator Award in 2024, and the Hong Kong RGC Research Fellowship
Scheme Award in 2024.
\end{IEEEbiography}

\begin{IEEEbiography}[{\includegraphics[width=1in,height=1.25in,clip]{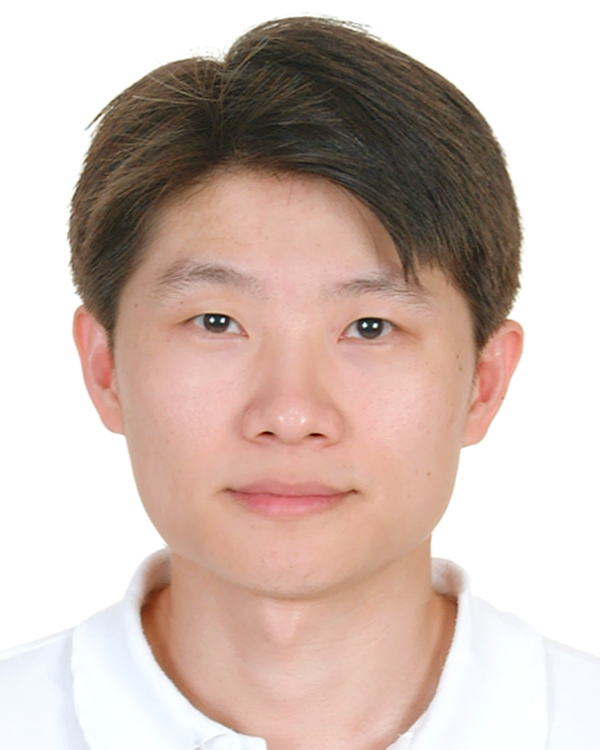}}]{Chun-Yao Wang}
(Senior Member, IEEE) received the Ph.D. degree from the Department of
Electronics Engineering, National Chiao Tung University, Hsinchu, Taiwan, in 2002.  He
joined the Department of Computer Science, National Tsing Hua University,
Hsinchu, in 2003 and is currently a Distinguished Professor.  His research
interests include logic synthesis, optimization, and verification for very
large-scale integrated and system-on-chip designs and emerging technologies.
\end{IEEEbiography}
\endgroup

\end{document}